\documentclass[aps,prd,superscriptaddress,eqsecnum]{revtex4}
\usepackage{graphicx}
\usepackage{color}
\usepackage{amssymb}
\begin{document}

\title{Inhomogeneous Condensates in the Thermodynamics  of the Chiral ${\rm NJL}_2$ model}
\author{G\"ok\c ce Ba\c sar}
\affiliation{Physics Department, University of Connecticut, Storrs CT 06269, USA}
\author{Gerald V. Dunne}
\affiliation{Physics Department, University of Connecticut, Storrs CT 06269, USA}
\author{Michael Thies}
\affiliation{Institut f\"ur  Theoretische Physik, Universit\"at Erlangen-N\"urnberg, D-91058,
Erlangen, Germany }

\begin{abstract}
We analyze the thermodynamical properties, at finite density and nonzero temperature, of the
 $(1+1)$-dimensional chiral Gross-Neveu model (the ${\rm NJL}_2$ model), using the exact inhomogeneous (crystalline) condensate solutions to the gap equation. The continuous chiral symmetry of the model plays a crucial role, and the thermodynamics leads to a broken phase with a periodic spiral condensate, the ``chiral spiral'', as a thermodynamically preferred  limit of the more 
general ``twisted kink crystal''  solution of the gap equation. This situation should be 
contrasted with the Gross-Neveu model, which has a discrete chiral symmetry, and for which 
the phase diagram has a crystalline phase with a periodic kink crystal.  We use a combination of analytic, numerical and Ginzburg-Landau techniques to study various parts of the phase diagram.

\end{abstract}
                                          
\maketitle

\section{Introduction}

The phase diagram of interacting fermion systems at finite density and temperature  is a general problem with applications in a wide range of physical contexts. Well-studied examples include the Peierls-Frohlich model of conduction \cite{peierls}, the Gorkov-Bogoliubov-de Gennes approach to superconductivity \cite{degennes}, and the Nambu-Jona Lasinio (NJL) model of symmetry breaking in particle physics \cite{nambu}.  Strongly interacting theories such as quantum chromodynamics (QCD) exhibit a rich phase diagram structure \cite{rajagopal,alford,casalbuoni}. It is known that chiral symmetry plays a key role, and computationally the large $N_f$ and large $N_c$ limits must be addressed carefully \cite{mclerran,glozman}.
A $(1+1)$-dimensional version of the NJL model, the ${\rm NJL}_2$ model [also known as the {\it chiral} Gross-Neveu model, $\chi{\rm GN}_2$] is of interest because it captures some important features of QCD, such as  asymptotic freedom, dynamical mass generation, a large $N_f$ limit, and the breaking of a continuous chiral symmetry  \cite{gross,dhn, shei,fz}. In this paper we use the exact crystalline solutions to the associated gap equation, found recently in  \cite{bd1,bd2}, to study the  temperature-density phase diagram of this ${\rm NJL}_2$ system.  The result of our thermodynamical analysis confirms the physical picture proposed in \cite{schon} that there is a phase transition at a critical temperature $T_c$ from a massless phase to a broken phase with a helical  condensate (the ``chiral spiral''), of the complex Larkin-Ovchinikov-Fulde-Ferrell (LOFF) form. The resulting phase diagram [see below, Figure \ref{fig5}], is very different from that of the non-chiral Gross-Neveu (${\rm GN}_2$) model, which has just a discrete, rather than continuous, chiral symmetry. In the ${\rm GN}_2$ model there is also a region of the phase diagram with a crystalline order parameter \cite{thies-gn}, but the structure of the phase diagram is very different [see below, Figure \ref{fig7}]. This crystalline phase of ${\rm GN}_2$ has been clearly seen in a recent lattice analysis \cite{deforcrand}, extending an important earlier lattice analysis \cite{karsch}. In this paper we explain in detail the role of the chiral symmetry [continuous versus discrete] in determining the form of the phase diagram. The chiral spiral phase of the ${\rm NJL}_2$ model has also been studied in the AdS/QCD framework \cite{davis}.
These one dimensional models are of course simplified models of more realistic $(3+1)$-dimensional systems, but important lessons can still be learned concerning the appearance of crystalline structures in the phase diagram \cite{bringoltz,nickel}. Furthermore, their solubility permits a detailed study of the relation between real and imaginary chemical potential \cite{thies-mu}. 

Our analysis is ultimately based on solving the gap equation for inhomogeneous condensates. Initially, the phase diagram of the ${\rm NJL}_2$ and ${\rm GN}_2$ models was studied assuming homogeneous condensates \cite{wolff,treml}, but this assumption does not capture certain aspects of the true physical phase diagram \cite{karsch,thies-gn,deforcrand}. Of course, finding inhomogeneous solutions to the gap equation is a much more difficult technical problem, but the massless ${\rm NJL}_2$ and ${\rm GN}_2$ models have remarkable symmetry properties that enable one to find the general periodic condensate solutions \cite{bd1,bd2}. This fact is due to a  deep connection between the Bogoliubov-de Gennes effective hamiltonian of the ${\rm NJL}_2$ system, and certain one dimensional integrable hierarchies \cite{pap,gesztesy,correa}. These exact solutions are characterized by a finite number of parameters, and to describe the phase diagram one must minimize the thermodynamical grand potential with respect to these parameters in order to determine the form of the condensate in a given region of the $(T, \mu)$ plane. This thermodynamical analysis is performed in this paper.

In Section II we briefly review the analytical solution of the inhomogeneous gap equation. In Section III we identify the special role played by rescaling and phase rotation symmetries in the ${\rm NJL}_2$ model. The thermodynamics of the ${\rm NJL}_2$ model is discussed in terms of a spiral condensate in Section IV and in terms of the general twisted kink crystal in Section V. In Section VI we contrast this analysis with the case of the ${\rm GN}_2$ model,  which has just a discrete chiral symmetry. In Section VII we apply a Ginzburg-Landau analysis to study the region of the phase diagrams of both the  ${\rm GN}_2$ and ${\rm NJL}_2$ models, in the vicinity of the relevant  ``tricritical point''. We conclude with a summary of our results and a discussion of the implications for more complicated models.

\section{Solving the Inhomogeneous Gap Equation}
\label{sec:gap}

The ${\rm NJL}_2$ model is described by the following $(1+1)$-dimensional  Lagrangian with both scalar 
and pseudoscalar four-fermion interaction terms:
\begin{equation}
{\mathcal L}_{\rm NJL}=\bar{\psi}\,i\, \partial \hskip -6pt / \,\psi +\frac{g^2}{2}\left[\left(\bar{\psi}\psi\right)^2
+\left(\bar{\psi}i\gamma^5 \psi\right)^2\right]\quad .
\label{lag-njl}
\end{equation}
This model has a continuous chiral symmetry: $\psi\to e^{i\gamma^5 \alpha} \psi$. The ${\rm GN}_2$ model has just the scalar four-fermion interaction term:
\begin{equation}
{\mathcal L}_{\rm GN}=\bar{\psi}\,i\, \partial \hskip -6pt / \,\psi +\frac{g^2}{2}\left[\left(\bar{\psi}\psi\right)^2
\right]\quad ,
\label{lag-gn}
\end{equation}
and has a discrete chiral symmetry: $\psi\to \gamma^5 \psi$. We study these models in the large $N_f$  limit where the semiclassical approximation applies and chiral symmetry breaking can be studied \cite{witten, affleck}.

By a Hubbard-Stratonovich transformation, the four-fermion interaction terms can be expressed in terms of
 scalar and pseudo-scalar bosonic condensate fields, $\Sigma$ and $\Pi$ (respectively), which are
 conveniently expressed in terms of a complex condensate field: $\Delta=\Sigma-i\Pi$. For ${\rm GN}_2$ we only have $\Sigma$, and so the condensate field $\Delta$ is real. The general  ${\rm NJL}_2$ system
 can be described equivalently by the effective Lagrangian:
\begin{equation}
{\mathcal L}=\bar{\psi}\left[\,i\, \partial \hskip -6pt / \,-\frac{1}{2}(1-\gamma^5)\Delta-\frac{1}{2}
(1+\gamma^5)\Delta^*\right]\psi -\frac{1}{2g^2}|\Delta|^2 ,
\label{lag2}
\end{equation}
which is now quadratic in the fermion fields. The corresponding single particle fermionic Hamiltoninan is \begin{equation}
H=- i  \gamma^5\frac{d}{dx}+\gamma^0\left(\frac{1}{2}(1-\gamma^5)\Delta-\frac{1}{2}(1+\gamma^5)\Delta^*\right)
\label{ham1}
\end{equation}
With the choice of the Dirac matrices as $\gamma_0=\sigma_1$, $\gamma_1=-i\sigma_2$ and $\gamma^5=\sigma_3$,
 the Hamiltonian (\ref{ham1}) takes the form:
\begin{equation}
H
=
\begin{pmatrix}
{-i\frac{d}{dx}&\Delta(x)\cr \Delta^*(x) & i\frac{d}{dx}}
\end{pmatrix}
\label{ham2}
\end{equation}
This Hamiltonian is also known as the Bogoliubov-de Gennes (BdG) (or Andreev)  Hamiltonian in the
 superconductivity literature \cite{degennes,leggett}.

There are two equivalent perspectives on studying the semiclassical gap equation for static condensates.
 The first, a Hartree-Fock approach, is to solve the single particle equation (the Bogoliubov-de Gennes
 equation)
\begin{equation}
H\psi=E\psi
\label{bdg}
\end{equation}
subject to the consistency condition relating the condensate field to the expectation values of the scalar 
and pseudoscalar fermionic bilinears:
\begin{eqnarray}
\langle\bar{\psi}\psi\rangle-i \langle\bar{\psi} i \gamma^5\psi\rangle=-\Delta/g^2 
\label{hf-condition}
\end{eqnarray}
A second approach is to integrate out the fermionic field in (\ref{lag2}) and obtain an effective action (per fermion flavor) for
 the condensate field:
\begin{equation}
S_{\rm eff}[\Delta]=-\frac{1}{2g^2 N_f}\int d^2x|\Delta|^2-i \ln\det\left[\,i\, \partial \hskip -6pt / 
\,-\frac{1}{2}(1-\gamma^5)\Delta-\frac{1}{2}(1+\gamma^5)\Delta^*\right]
\label{s_eff}
\end{equation}
The gap equation for the condensate field is obtained by looking for the stationary points of $S_{\rm eff}
[\Delta]$:
\begin{eqnarray}
0&=&\frac{\delta S_{\rm eff}}{\delta\Delta^*}\nonumber\\
&=&-\frac{1}{2g^2 N_f}\Delta(x)-i\frac{\delta}{\delta\Delta(x)^*}\ln\det\left[\,i\, \partial \hskip -6pt /
 \,-\frac{1}{2}(1-\gamma^5)\Delta(x)-\frac{1}{2}(1+\gamma^5)\Delta(x)^*\right]
\label{gap1}
\end{eqnarray}
It is straightforward to solve this gap equation when the condensate field $\Delta$ is uniform, but it is
 more technically challenging to solve it for an inhomogeneous condensate field $\Delta(x)$. Nevertheless, 
in one spatial dimension it is possible to find the most general bounded quasi-periodic solution to this 
gap equation \cite{bd1,bd2}. The general solution has the form of a ``twisted kink crystal'', described below.

A useful quantity for solving the inhomogeneous gap equation (\ref{gap1}) is the resolvent $R(x; E)$, 
 the coincident-point limit of the Gor'kov Green's function $G(x, y; E)$ corresponding to the Hamiltonian 
(\ref{ham2})
\begin{equation}
R(x; E)\equiv \langle x| \frac{1}{H-E} |x\rangle \quad .
\label{res}
\end{equation}
For a \textit{static} condensate the gap equation (\ref{gap1}) can be written as 
\begin{eqnarray}
\Delta(x)=-i N_f g^2\, {\rm tr}_{D,E}\left[\gamma^0\left({\bf 1}+\gamma^5\right)R(x; E)\right]
\label{gap2}
\end{eqnarray}
The solution of the gap equation relies on the remarkable fact that in one spatial dimension the resolvent 
(itself a $2\times 2$ matrix) must satisfy a simple first order matrix differential equation
\begin{eqnarray}
\frac{\partial}{\partial x}R(x;E)\sigma_3=i[
\begin{pmatrix}
{E&-\Delta(x)\cr \Delta^*(x) &-E}
\end{pmatrix}
, R \sigma_3]
\label{dikii}
\end{eqnarray}
This equation is known as the Eilenberger equation in the superconductivity literature \cite{eilenberger, 
leggett}, and as the Dickey equation in mathematical physics \cite{dickey,gesztesy}.  The 
Dickey-Eilenberger equation follows immediately from the fact that the resolvent can be written as a 
product of two linearly independent solutions:
\begin{eqnarray}
R(x; E)=\frac{1}{2i\, W}\left(\psi_1\psi_2^T+\psi_2\psi_1^T\right) \sigma_1
\label{product}
\end{eqnarray}
where $W=i(\psi_1^T\sigma_2\psi_2)$ is the Wronskian of  two independent solutions $\psi_{1,2}$ of 
$H\psi=E\psi$.

The inhomogeneous gap equation (\ref{gap1}) can be solved by the following simple ansatz \cite{bd1,bd2} for the resolvent
\begin{eqnarray}
R(x; E)={\mathcal N}(E)
\begin{pmatrix}
{a(E)+|\Delta|^2&b(E)\,\Delta-i\Delta^\prime \cr b(E)\,\Delta^*+i\Delta^{*\,\prime} &a(E)+|\Delta|^2}
\end{pmatrix}
\label{ansatz}
\end{eqnarray}
where ${\mathcal N}(E)$, $a(E)$ and $b(E)$ are functions of the energy $E$, and are to be determined.
This particular ansatz is motivated by the gap equation (\ref{gap2}) that relates the off-diagonal component
 of the resolvent with $\Delta$. The ansatz (\ref{ansatz}) automatically solves the diagonal part of  the
 Eilenberger equation (\ref{dikii}), while the off-diagonal part requires that the condensate field $\Delta$
 satisfy the complex nonlinear Scr\"odinger equation (NLSE):
\begin{equation}
\Delta^{\prime\prime} -2|\Delta |^2\,\Delta +i\left(b(E)-2 E\right)\Delta^\prime -2\left(a(E)-E\, b(E)\right)
\Delta=0
\label{nlse}
\end{equation}
The advantage of this ansatz approach is that the NLSE (\ref{nlse}) can be solved in closed form, and its 
general solution has the form of a twisted kink crystal, described in detail in \cite{bd2} and summarized 
below in the next section. The associated energy functions ${\mathcal N}(E)$, $a(E)$ and $b(E)$ are simple 
functions of $E$. For the ${\rm NJL}_2$ model, there is a further consistency condition 
required to satisfy the gap equation (\ref{gap2}): for an inhomogeneous condensate, the part of the 
off-diagonal resolvent proportional to $\Delta^\prime(x)$ must vanish. This places a condition on the 
energy function ${\mathcal N}(E)$:
\begin{eqnarray}
0={\rm tr}_E\left({\mathcal N}(E)\right)\equiv \int \frac{dE}{2\pi} \frac{{\mathcal N}(E)}{1+e^{\beta(E-\mu)}}
\label{cc}
\end{eqnarray}
Here $\beta=1/T$ is the inverse temperature and $\mu$ is the chemical potential.
This consistency condition imposes one relation on the parameters describing the twisted kink solution. With this 
consistency condition imposed,  the general inhomogeneous condensate $\Delta(x)$ satisfying the NLSE (\ref{nlse}) solves the gap 
equation (\ref{gap1}). 

Given this exact solution $\Delta(x)$ to the gap equation (\ref{gap1}), it is also possible to find the 
exact single-particle solutions to the BdG equation (\ref{bdg}). Furthermore, the diagonal resolvent $R(x; E)$  in (\ref{ansatz}) encodes all the 
relevant spectral information. For example, the local density of states for fermions in the presence of the 
condensate is given by
\begin{eqnarray}
\rho(x;E)=\frac{1}{\pi}{\rm Im}\,{\rm tr}_D\left(R(x; E+i\epsilon)\right) 
\label{local_dos}
\end{eqnarray}
where the matrix trace of the resolvent follows trivially from the ansatz (\ref{ansatz}):
\begin{eqnarray}
{\rm tr}_D\left(R(x; E)\right) =2{\mathcal N}(E)\left(a(E)+|\Delta(x)|^2\right)
\label{trace}
\end{eqnarray}
Given the density of states $\rho(E)=\int dx\, \rho(x; E)$, all relevant thermodynamic quantities, at 
finite temperature and chemical potential, can be derived from the grand canonical potential
\begin{equation}
\Psi[\Delta(x); T, \mu]= -\frac{1}{\beta}\int_{-\infty}^{\infty}dE\,\rho(E)\ln\left(1+e^{-\beta(E-\mu)}
\right) + \frac{1}{2{N_f}g^2} \frac{1}{L} \int_0^L dx |\Delta(x)|^2 \quad .
\label{grand}
\end{equation}
 Since we know $\rho(E)$ exactly, we can analyze the thermodynamical properties of this
 model precisely. 

\subsection{Twisted Kink Crystal Condensate}
\label{sec:gap-twisted}

\begin{figure}[h]
\includegraphics[scale=0.75]{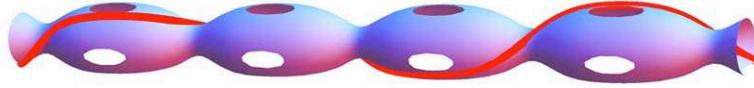}
\caption{The twisted kink crystal condensate of (\ref{complex-kink-crystal}), shown as the red curve. The blue skeleton surface is shown just to illustrate the  periodic amplitude modulation and phase winding.} 
\label{fig1}
\end{figure}

The general solution to the NLSE (\ref{nlse}) describes a crystalline condensate \cite{bd1,bd2}. It is a
 periodic array of kinks that also rotate in the chiral plane, as illustrated in Figure \ref{fig1}. The
 {\it single} chirally-twisted kink was originally found by Shei \cite{shei} using inverse scattering techniques, 
and subsequently studied in a resolvent approach by Feinberg and Zee \cite{fz}. The periodic array 
of such twisted kinks can be expressed in terms of the elliptic functions:
\begin{eqnarray}
\Delta(x)&=&-\lambda\, e^{2i q x}\, A\,\frac{\sigma(\lambda A\, x+i{\bf K}^\prime -i\theta/2)}
{\sigma(\lambda A\, x+i{\bf K}^\prime)\sigma(i\theta/2)}
\,\exp\left[i\lambda A\, x \left(-i\,\zeta(i\theta/2)+i\,{\rm ns}(i\theta/2)\right)+i\,\theta \eta_3/2\right]
 \quad 
\label{complex-kink-crystal}
\end{eqnarray}
where sc=sn/cn, nd=1/dn are Jacobi elliptic functions, and the functions $\sigma$ and $\zeta$ are the
 Weierstrass sigma and zeta functions  \cite{lawden}, chosen to have real and imaginary half-periods: $\omega_1={\bf K}
(\nu)$, and $\omega_3=i\,{\bf K}^\prime\equiv i{\bf K}(1-\nu)$. 
Both periods are therefore controlled by a single [real] elliptic parameter $0\leq \nu\leq 1$. Note that 
$\eta_3=\zeta(i{\bf K}^\prime)$ is pure imaginary. The parameter $\lambda$ sets the overall scale of the 
condensate, and $1/\lambda$ sets the length scale of the crystal. 
Later, we will use units in which the vacuum mass of the fermion is $1$, so that 
$\lambda$ sets the scale relative to the vacuum fermion mass.
The angular parameter $\theta$ takes 
values in the range $\theta\in [0, 4{\bf K}^\prime(\nu)]$. The [real] constant $A$ is  a function of 
$\theta$ and the elliptic parameter $\nu$:
\begin{eqnarray}
A=A(\theta,\nu)=-2i\,{\rm sc}(i\theta/4; \nu)\,{\rm nd}(i\theta/4; \nu) 
\label{a}
\end{eqnarray}
For brevity we will usually suppress the explicit dependence of the elliptic functions on the elliptic parameter $\nu$.
The final parameter $q$ is a phase parameter that affects the chiral angle through which the condensate 
rotates over one period $L=\frac{2{\bf K}}{\lambda A}$ : 
\begin{equation}
\Delta(x+L)=e^{2i\varphi} \Delta(x) \quad ; \quad \varphi= {\bf K}\left(-i\,\zeta(i\theta/2)+i\,{\rm ns}
(\theta/2)-\frac{\eta \theta}{2{\bf K}}+\frac{2q}{\lambda A}\right)
\label{winding}
\end{equation}
where $\eta\equiv \zeta({\bf K})$  is real.
Thus the general solution is specified by four real parameters: a scale parameter $\lambda$, a phase 
parameter $q$, an angular parameter $\theta$, and the elliptic parameter $\nu$. These parameters also 
parametrize the energy spectrum of fermions in such a condensate background, which has two gaps, with 
band edges  $E_1\leq E_2\leq E_3\leq E_4$ as shown in the first plot of Figure \ref{fig2}: 
 \begin{figure}[h]
\centerline{\includegraphics[scale=0.35]{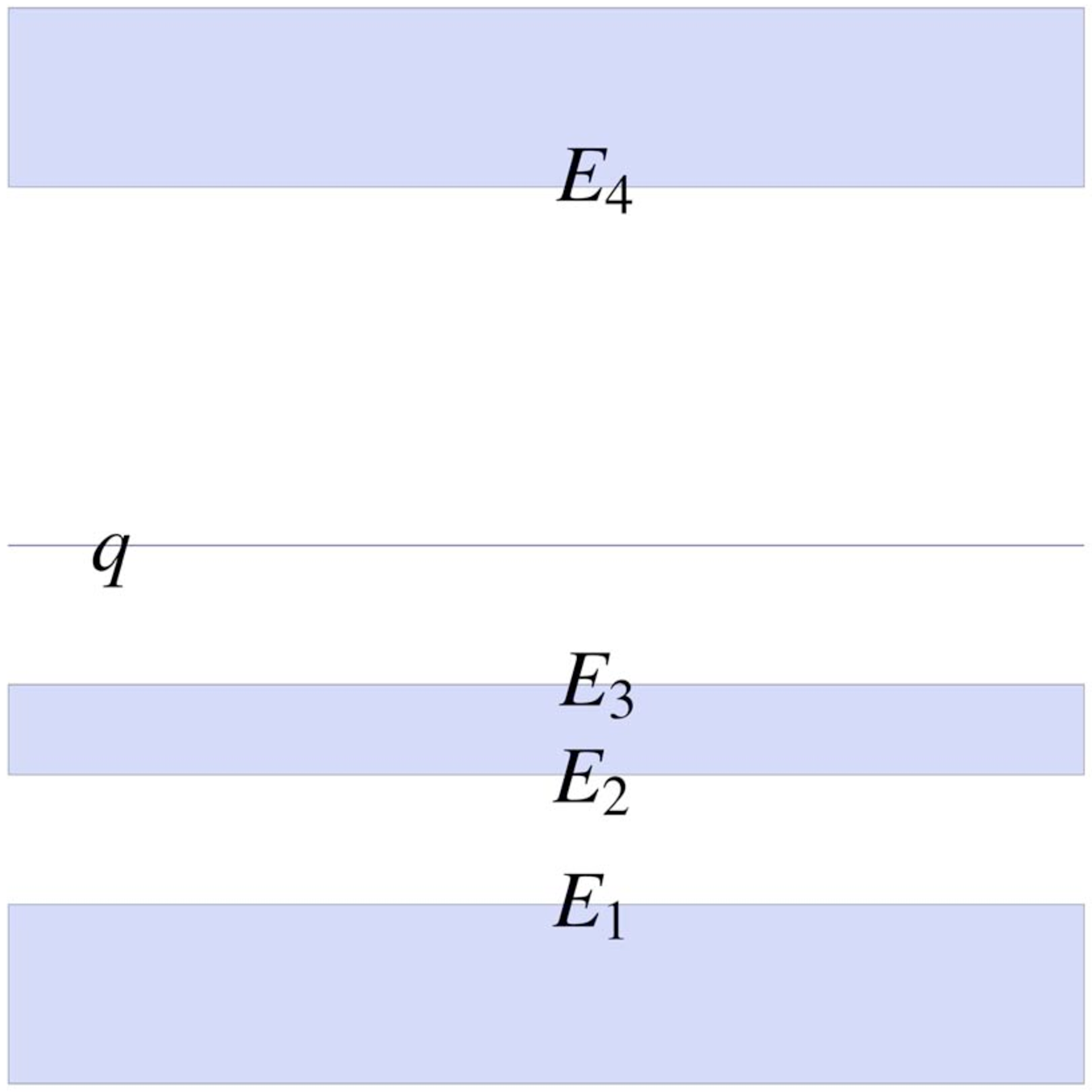}
\includegraphics[scale=0.35]{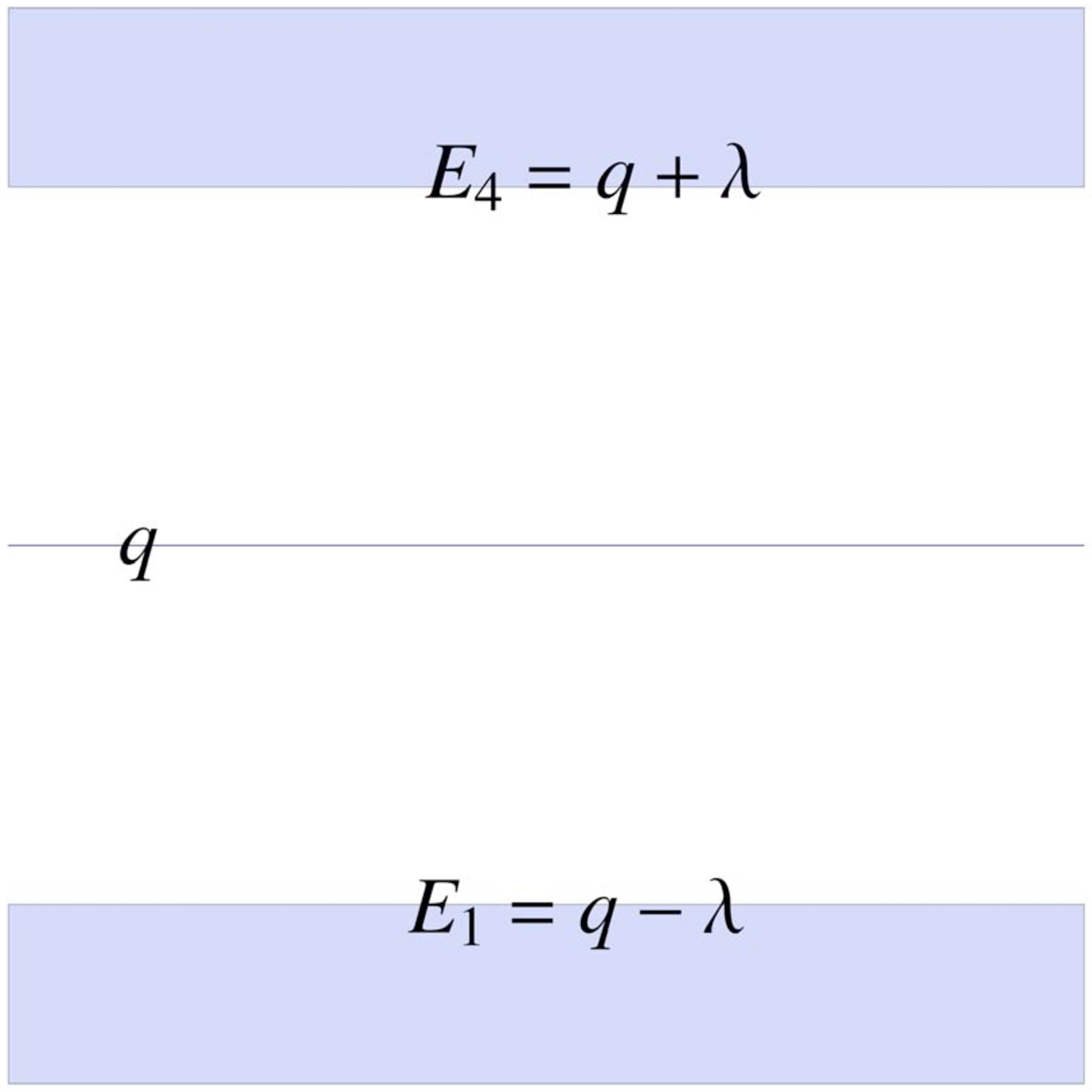}
\includegraphics[scale=0.35]{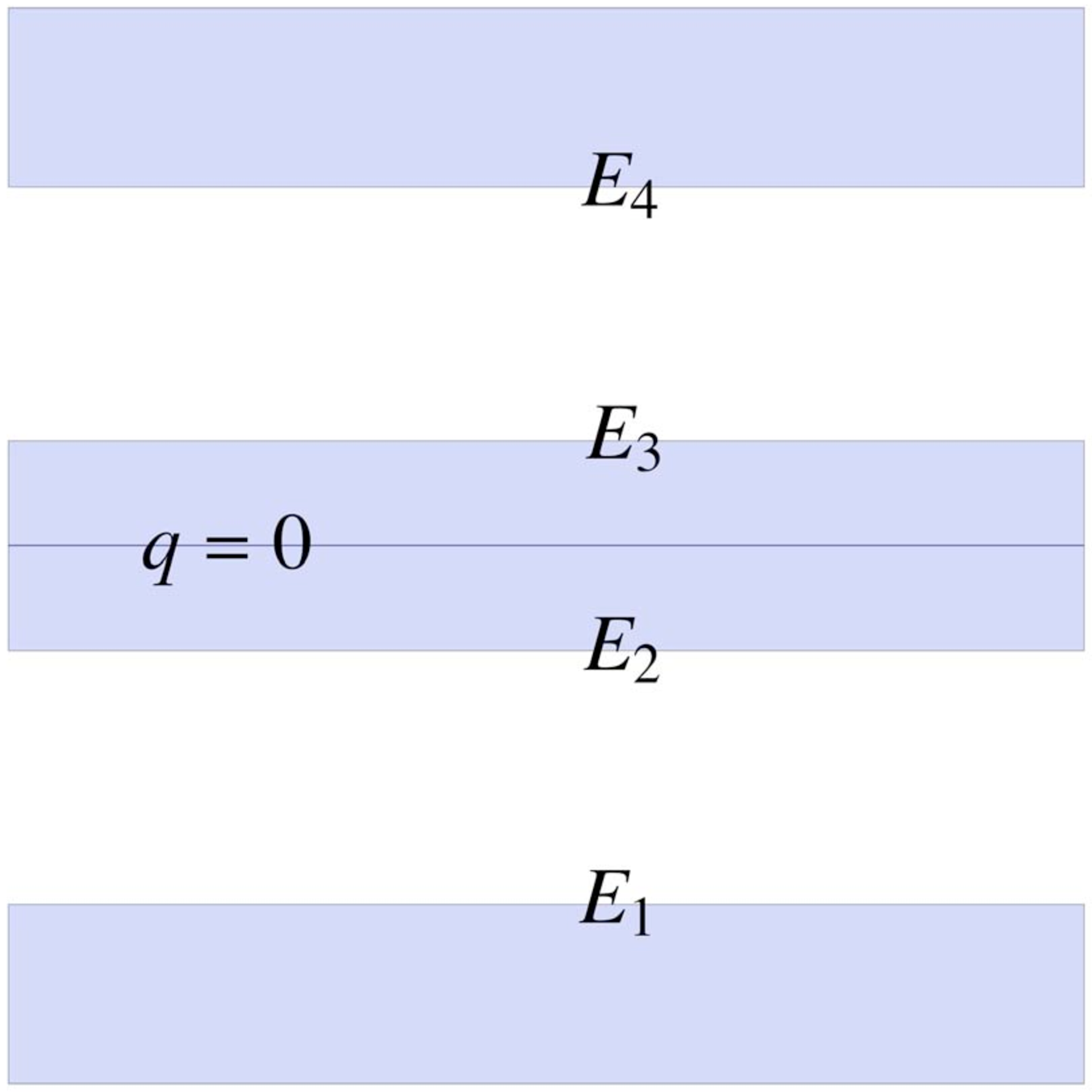}}
\caption{The form of the single-particle fermion spectra for the general twisted kink crystal [first figure], showing the central value $E=q$, and the band edges $E_j$, for $j=1\dots 4$. The second figure shows the special case of the spiral condensate, for which the bound band merges with one of the continua. The third figure shows the spectrum for another special case, the real kink crystal, which has a charge conjugation symmetry, implying that the offset is $q=0$, and the spectrum is symmetric about 0. The position of the bands within the gap, and their width, are controlled by the parameters $\theta$ and $\nu$.}
\label{fig2} 
\end{figure}
\begin{eqnarray}
E_1&=&q-\lambda
\nonumber \\
E_2&=&q+\lambda(-1+2\,{\rm nc}^2(i\theta/4))
\nonumber \\
E_3&=&q+\lambda(-1+2\,{\rm nd}^2(i\theta/4))
\nonumber \\
E_4&=&q+\lambda
\label{band_edges}
\end{eqnarray}
Thus, in terms of the single particle fermion spectrum, the role of the four parameters is as follows:
 $\lambda$ determines the overall energy scale; $q$ determines the overall offset; while $\theta$ and 
$\nu$ determine the location and width of the band that lies in the gap between the ``outer'' edges $E_1$ 
and $E_4$. The simple linear dependence of the energy spectrum on the parameters $\lambda$ and $q$ is a direct consequence of 
the form of the Hamiltonian (\ref{ham2}), and reflects the an important scale and shift symmetry described in detail in Section \ref{sec:scale}.

For the twisted kink crystal solution (\ref{complex-kink-crystal}), the coefficients $a(E)$ and $b(E)$ in the NLSE (\ref{nlse}) are simple polynomials of $E$, with coefficients determined by the band edges:
\begin{eqnarray}
a(E)&=&2E^2-\left(\sum_{j=1}^4 E_j\right) E+\frac{1}{8}\left\{ \left(\sum_{j=1}^4 E_j\right) ^2-\sum_{i<j}^4 
(E_i-E_j)^2\right\} \\
b(E)&=&2 E-\left(\sum_{j=1}^4 E_j\right) 
\label{ab-njl}
\end{eqnarray}
Furthermore, the energy function ${\mathcal N}(E)$ appearing in the resolvent ansatz (\ref{ansatz})
 also has a very simple form in terms of the band edges: 
\begin{eqnarray}
{\mathcal N}(E)=\frac{i}{4\sqrt{\prod_{j=1}^4(E-E_j)}}
\label{n-njl}
\end{eqnarray}
Thus, we have an explicit exact expression for the density of states of fermions in the presence of such
 a twisted kink condensate field, following from the trace of the resolvent. Within the bands: 
  \begin{eqnarray}
\rho(E)=\frac{1}{2\pi} \frac{a(E)+\lambda^2Z}{\sqrt{\prod_{j=1}^4(E-E_j)}}
\label{dos}
\end{eqnarray}
Here we have defined the function $Z(\theta, \nu)$ in terms of the normalized average of $|\Delta(x)|^2$ over one period:
\begin{equation}
Z(\theta, \nu)\equiv \frac{1}{\lambda^2}\langle|\Delta(x)|^2\rangle=-A(\theta,\nu)^2\left({\mathcal P}
(i \theta/2)+\frac{\eta}{{\bf K}}\right)
\label{zed}
\end{equation}
with ${\mathcal P}$ being the Weierstrass ${\mathcal P}$ function. Thus, the density of states $\rho(E)$ 
is an explicitly known function of the energy $E$, depending parametrically on the four parameters $\lambda$, 
$q$, $\theta$ and $\nu$ that characterize the solution (\ref{complex-kink-crystal}) to the gap equation. This parametric dependence enters through the band edge energies $E_j$ in (\ref{band_edges}), and through the function $Z$ defined in (\ref{zed}).

\subsection{Spiral Condensate}
\label{sec:gap-spiral}

An important special case of the general solution (\ref{complex-kink-crystal}) is the degenerate case 
when the bound band of the fermion spectrum shrinks and merges with the upper or lower continuum, so 
that the spectrum has just a single gap, as shown in the second plot in Figure \ref{fig2}. This occurs when the angular
 parameter takes values at its extreme limits: $\theta=0$ [which implies that $E_2=E_3=E_4$, so that the bound band merges with the upper continuum], or  
$\theta=4 {\bf K}^\prime$  [which implies that $E_1=E_2=E_3$, so that the bound band merges with the lower continuum]. The general twisted kink crystal condensate
 (\ref{complex-kink-crystal}) reduces to a single plane wave
\begin{equation}
\Delta=\lambda \, e^{2iqx}
\label{complex-plane}
\end{equation}
\begin{figure}[h]
\includegraphics[scale=0.75]{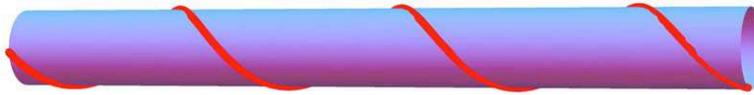}
\caption{The spiral condensate of (\ref{complex-plane}), shown as the red curve. The blue skeleton surface is shown just to illustrate the  periodic phase winding. In contrast to the twisted kink crystal in Figure \ref{fig1}, for the spiral, the amplitude is constant.} 
\label{fig3}
\end{figure}
which is clearly a solution to the NLSE (\ref{nlse}). For this condensate the amplitude is constant, while the phase 
rotates at a constant rate, set by $q$, as shown in Figure \ref{fig3}. The  fermion energy spectrum has 
just one gap, of width $2\lambda$, centered at $q$; that is, the band edges lie at $E_1=q-\lambda$, and 
$E_4=q+\lambda$. Correspondingly, the resolvent trace has a simplified form, and the spectral function within the continuum bands is simply:
\begin{eqnarray}
\rho( E)=\frac{1}{\pi} \frac{|E-q|}{\sqrt{\lambda^2-(E-q)^2}}
\end{eqnarray}
which we recognize as the spectral function of a constant condensate $\Delta=\lambda$, shifted in energy by $q$.

\subsection{Real Kink Crystal}
\label{sec:gap-real}

Another important special case of the general solution (\ref{complex-kink-crystal})  is the case where
 the condensate is real [relevant for the ${\rm GN}_2$ model], which implies that the  BdG Hamiltonian $H$ in (\ref{ham2}) has a charge-conjugation symmetry, 
$\{ H, \sigma_2\}=0$, which in turn implies that the fermionic spectrum is symmetric, as shown in the third plot of 
Figure \ref{fig2}. The band edges reduce to 
\begin{eqnarray}
E_1&=&-\lambda=-E_4
\nonumber \\
E_2&=&-\lambda\left(\frac{1-\sqrt{\nu}}{1+\sqrt{\nu}}\right)=-E_3
\label{gn_band_edges}
\end{eqnarray}
\begin{figure}[h]
\includegraphics[scale=0.75]{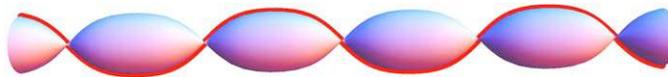}
\caption{The real kink crystal condensate of (\ref{real-kink-crystal}), shown as the red curve. The blue skeleton surface is shown just to illustrate the  periodic amplitude modulation and phase winding. For this real kink crystal, the amplitude vanishes each period, and the kink rotates through $\pi$ [i.e., changes sign] each period. } 
\label{fig4}
\end{figure}
The phase parameter $q=0$, and further the angular parameter $\theta$ takes its midpoint value 
$\theta=2{\bf K}^\prime(\nu)$. Thus the real kink crystal is described by just two parameters, 
the scale $\lambda$ and the elliptic parameter $\nu$:
\begin{eqnarray}
\Delta(x)&=&\lambda\left(\frac{2 \sqrt{\nu}}{1+\sqrt{\nu}}\right){\rm sn}\left( \frac{2\lambda\, x}
{1+\sqrt{\nu}}; \nu \right)\nonumber\\
&=&\lambda\, \tilde{\nu}\, \frac{{\rm sn}\left( \lambda\,x; \tilde{\nu} \right) {\rm cn}
\left( \lambda\,x; \tilde{\nu} \right)}{{\rm dn}\left( \lambda\,x; \tilde{\nu} \right)} 
\qquad ; \qquad \tilde{\nu}\equiv\frac{4\sqrt{\nu}}{(1+\sqrt{\nu})^2} \quad .
\label{real-kink-crystal}
\end{eqnarray}
The second form of $\Delta(x)$ in (\ref{real-kink-crystal}) is obtained from the first form by a Landen transformation \cite{lawden}.
Over one period, $L=\frac{2{\bf K}(\tilde{\nu})}{\lambda}$, the condensate changes sign [that is, it rotates through an angle $2\varphi=-\pi$], as shown in 
Figure \ref{fig4}. This change of sign corresponds to the {\it discrete} chiral symmetry of the 
${\rm GN}_2$ model, while the phase rotation (\ref{winding}) of the general kink crystal condensate
 (\ref{complex-kink-crystal}) is associated with the {\it continuous} chiral symmetry of the 
${\rm NJL}_2$ model. The real kink crystal describes the inhomogeneous condensate of the crystalline 
phase of the ${\rm GN}_2$ model \cite{thies-gn}, and its thermodynamics will be discussed below 
in Section {\ref{sec:gn}.

\section{The Scale and Phase Symmetry in ${\rm NJL}_2$}
\label{sec:scale}

 In this Section we describe   a simple but important symmetry 
property of the Bogoliubov-de Gennes equation (\ref{bdg}), that has important consequences for the 
thermodynamical analysis. The Bogoliubov-de Gennes equation (\ref{bdg}) admits a family of solutions
 obtained by rescaling and phase shifting (i.e. making a {\it linear} local chiral rotation) a given solution:
\begin{eqnarray}
\Delta(x)\rightarrow\lambda\,\Delta(\lambda\, x)\, e^{2iqx}
\nonumber \\
\psi(x)\to e^{iqx \gamma_5} \lambda^{1/2} \,\psi(\lambda\, x)
\label{trans}
\end{eqnarray}
which generates all the linear transformations acting on the energy spectrum:
\begin{equation}
E\rightarrow\lambda\, E+q \quad .
\label{energy-trans}
\end{equation}
In terms of the density of states, the effect of the transformation is: 
\begin{equation}
\rho(E)\rightarrow\rho\left(\frac{E-q}{\lambda}\right) \quad .
\label{rho-trans}
\end{equation}
The important physical implication  of this symmetry is that when minimizing the grand potential 
(\ref{grand}) with respect to the four parameters $\lambda$, $q$, $\theta$ and $\nu$, the minimization with
 respect to $\lambda$ and $q$ can  be done first. If the grand potential did not require renormalization, 
then the minimization with respect to $\lambda$ and $q$  would be trivial. In fact, we will show 
in the next section that even taking into account the renormalization, these symmetries greatly simplify 
the minimization with respect to $\lambda$ and $q$.

It is useful to define the  "unscaled" and "unshifted" spectrum to be the one with $\lambda=1$ 
and $q=0$, so that $E_1=-1$, and $E_4=1$  (in units where the vacuum fermion mass is 1).
All other spectral functions can be generated from this basic 
solution using the simple transformation (\ref{rho-trans}). The corresponding density of states will be written as 
\begin{equation}
\hat{\rho}(E)=\frac{1}{2\pi} \frac{\left(2E^2-(\hat{E}_2+\hat{E}_3)E-(\hat{E}_3-\hat{E}_2)^2/4-1+Z\right)}{ \sqrt{(E^2-1)(E-\hat{E}_2)(E-\hat{E}_3)}}
\label{bare-dos}
\end{equation}
where $Z=Z(\theta, \nu)$ is defined in (\ref{zed}), and
\begin{eqnarray}
\hat{E}_2&=&-1+2\,{\rm nc}^2(i\theta/4; \nu)
\nonumber \\
\hat{E}_3&=&-1+2\,{\rm nd}^2(i\theta/4; \nu)
\label{bare_band_edges}
\end{eqnarray}
Importantly, $\hat{\rho}(E)$  depends parametrically only on the two remaining parameters, $\theta$ and $\nu$. This separation of parametric dependences has important consequences for the minimization of the thermodynamic grand potential (\ref{grand}) with respect to the parameters.

\subsection{Transformation Properties of Thermodynamic Quantities}
 
\subsubsection{The Grand Potential $\Psi$}
We begin our discussion with the grand canonical potential $\Psi[\hat{\Delta}(x);T,\mu] $ 
 for the unscaled/unshifted condensate 
$\hat{\Delta}(x)$, obtained from (\ref{complex-kink-crystal}) by setting the scale parameter $\lambda=1$, and the phase parameter $q=0$. The grand potential is formally divergent in the UV region and has to be renormalized, as is well known \cite{gross,dhn,thies-gn}. At finite density and nonzero temperature, it is convenient  to separate the single particle contribution as 
\begin{equation}
-\frac{1}{\beta}\int_{-\infty}^{\infty}dE\, \hat{\rho}(E) \ln(1+e^{-\beta(E-\mu)}) 
= \int_{E_{\rm min}}^{\mu}dE\,  \hat{\rho}(E)(E-\mu)-\frac{1}
{\beta}\int_{-\infty}^{\infty}dE\, \hat{\rho}(E) \ln(1+e^{-\beta|E-\mu|})
\label{split}
\end{equation}
where $E_{\rm min}= -\Lambda/2-Z/\Lambda+\dots$, in terms of the  momentum cutoff $\Lambda/2$.
Only the first term, the zero temperature expression, in (\ref{split}) is divergent. We  isolate the 
divergent terms using the large $E$ behaviour of the density of states (\ref{bare-dos}):
\begin{equation}
 \hat{\rho}(E)\approx1+\frac{Z}{2E^2}+\dots
\end{equation}
The divergent part is
\begin{equation}
\Psi_{\rm div}=-\frac{\Lambda^2}{8\pi}-\frac{\Lambda\mu}{2\pi}-\frac{Z}{2\pi}\ln\Lambda \quad .
\end{equation}
The quadratically and linearly divergent terms are absorbed by definition of the renormalized energy and baryon 
number densities, and the logarithmically divergent term is canceled by the double counting correction \cite{thies-gn}
\begin{equation}
\frac{1}{2{N_f}g^2}\frac{1}{L}\int_0^L | \hat{\Delta}(x)|^2dx=\frac{Z}{2\pi} \ln\Lambda
\end{equation}
where we have used vacuum gap equation 
$
\frac{\pi}{{N_f}g^2}=\ln\Lambda
$.
Hence the finite renormalized grand canonical potential is
\begin{equation}
\Psi_{\rm ren}[\hat{\Delta}(x);T,\mu] = \int_{E_{\rm min}}^{\mu}dE\, \hat{\rho}(E)(E-\mu)-\frac{1}
{\beta}\int_{-\infty}^{\infty}dE\, \hat{\rho}(E)\ln(1+e^{-\beta|E-\mu|})
+\frac{\Lambda^2}{8\pi}+\frac{\Lambda\mu}{2\pi}+\frac{Z}{2\pi}\ln\Lambda
\label{psi_ren}
\end{equation}
Now we can analyse the effect of the transformation (\ref{rho-trans}) on the {\it renormalized} grand canonical potential  for the general condensate 
\begin{eqnarray}
\Delta(x)=\lambda\,\hat{\Delta}(\lambda\, x)\,e^{2iqx} \quad .
\label{general-delta}
\end{eqnarray}

The finite temperature (f.t.) contribution [the 2nd term on the r.h.s. of Eq.~(\ref{psi_ren})] has the following simple scaling behaviour,
\begin{eqnarray}
\left. \Psi_{\rm ren} [\lambda\,\hat{\Delta}(\lambda x)\,e^{2iqx}; T, \mu]\right|_{\rm f.t.}&=& 
-\frac{1}{\beta}\int_{-\infty}^{\infty}dE\,\hat{\rho}\left(\frac{E-q}{\lambda}\right) 
\ln(1+e^{-\beta|E-\mu|})
\nonumber \\
& = & - \frac{\lambda^2}{\hat{\beta}} \int_{-\infty}^{\infty} dE\, \hat{\rho}(E) \ln(1+e^{-\hat{\beta}|E-\hat{\mu}|})
\nonumber \\
& = & \left. \lambda^2\, \Psi_{\rm ren}[\hat{\Delta}(x); \hat{T}, \hat{\mu}]\right|_{\rm f.t.}
\label{Z1}
\end{eqnarray}
with the rescaled variables
\begin{eqnarray}
\hat{\mu}& = & \frac{\mu-q}{\lambda} 
\nonumber \\
\hat{\beta} & = & \frac{1}{\hat{T}} \ =\ \lambda \beta
\label{Z2}
\end{eqnarray}
For the zero temperature contribution (z.t.) in (\ref{psi_ren}), we start from the expression,
\begin{equation}
\left. \Psi_{\rm ren}[\lambda\,\hat{\Delta}(\lambda x)\,e^{2iqx}; T, \mu]\right|_{\rm z.t.} =  \int_{E_{\rm min}^{\lambda}}^{\mu}dE\,
 \hat{\rho}\left(\frac{E-q}{\lambda}\right)(E-\mu) +\frac{\Lambda^2}{8\pi}+\frac{\Lambda\mu}{2\pi}+\frac{\lambda^2 Z}{2\pi}\ln\Lambda
\label{Z3}
\end{equation}
where $E_{\rm min}^{\lambda}=-\Lambda/2-\lambda^2 Z/\Lambda$. Here, due to the regularization, the scaling relation analoguous
to Eq.~(\ref{Z1}) develops anomalous terms akin to the chiral U(1) and scale anomalies,
\begin{equation}
\left. \Psi_{\rm ren} [\lambda\,\hat{\Delta}(\lambda x)\,e^{2iqx}; T, \mu]\right|_{\rm z.t.} = \lambda^2 \left. \Psi_{\rm ren}[\hat{\Delta}(x);\hat{T}, \hat{\mu}]\right|_{\rm z.t.}
+ \frac{Z}{2\pi} \lambda^2 \ln \lambda + \lambda^2 \frac{\hat{\mu}^2}{2\pi} - \frac{\mu^2}{2\pi}
\label{Z4}
\end{equation} 
Being an UV effect, the extra terms are independent of temperature. Combining Eqs.~(\ref{Z1}) and (\ref{Z4}),
we see that the renormalized grand potential for the general condensate  in (\ref{general-delta}), is
\begin{equation}
\Psi_{\rm ren}[\lambda\, \hat{\Delta}(\lambda x)\, e^{2iqx}; T, \mu] = \lambda^2\left(\hat{\Psi}_{\rm ren}
+\frac{Z}{2\pi}\ln\lambda+\frac{\hat{\mu}^2}{2\pi}\right)-\frac{\mu^2}{2\pi}
\label{psi_trans}
\end{equation}
with the shorthand notation
\begin{equation}
\hat{\Psi}_{\rm ren} \equiv  \Psi_{\rm ren}[\hat{\Delta}(x);\hat{T},\hat{\mu}].
\label{Z5}
\end{equation}
For the sake of compactness in the notation, we will drop the subscript ``ren'' from now on, and work 
exclusively with the physical renormalized thermodynamic quantities.

The grand canonical potential is related to the density $\rho$, [not to be confused with the density 
of states $\rho(E)$!], the entropy $s$, and the free energy $u$:
\begin{equation}
\Psi=u-\mu\rho-Ts
\label{grand_canonical}
\end{equation}
Thus we can obtain expressions for the effect of the scaling and phase shifting transformation on the 
renormalized  $\rho$, $s$  and $u$ as follows:

\subsubsection{Number Density}
From the basic relation $\rho=-\frac{\partial\Psi}{\partial\mu}$,
we write $\frac{\partial}{\partial\mu}=\frac{1}{\lambda}\frac{\partial}{\partial\hat\mu}$, and act on 
(\ref{psi_trans}) to obtain:
\begin{eqnarray}
\rho&=&\lambda^2(-\frac{1}{\lambda}\frac{\partial\hat{\Psi}}{\partial\hat{\mu}}-\frac{1}{\lambda}
\frac{\hat\mu}{\pi})+\frac{\mu}{\pi}\nonumber\\
&=&\lambda\,{\hat\rho}+\frac{q}{\pi}
\label{density_trans}
\end{eqnarray}

\subsubsection{Entropy}
From the basic relation $s=-\frac{\partial\Psi}{\partial T}$,
we write $\frac{\partial}{\partial T}=\frac{1}{\lambda}\frac{\partial}{\partial \hat{T}}$, and act on 
(\ref{psi_trans}) to obtain:
\begin{eqnarray}
s&=&\lambda^2(-\frac{1}{\lambda}\frac{\partial\hat{\Psi}}{\partial\hat{T}})\nonumber\\
&=&\lambda\,\hat{s}
\label{entropy_density_trans}
\end{eqnarray}

\subsubsection{Free Energy}
The transformation property of the free energy now follows directly from the relation (\ref{grand_canonical}):
\begin{eqnarray}
u&=&\Psi+\mu\rho+Ts\nonumber\\
&=&\lambda^2\left(\hat{u}+\frac{Z}{2\pi}\ln\lambda\right)+\lambda\, q\, \hat\rho+\frac{q^2}{2\pi}
\label{energy_density_trans}
\end{eqnarray}

\subsection{Implications for Minimization of the Grand Potential $\Psi$ with respect to the phase 
parameter $q$}

The minimization of $\Psi$ with respect to $q$ can be transformed into minimization with respect to 
the chemical potential, due to the symmetry (\ref{rho-trans}). We write
 $\frac{\partial}{\partial q}=-\frac{1}{\lambda}\frac{\partial}{\partial \hat \mu}$, and  
differentiate $\Psi$ in (\ref{psi_trans}) with respect to $\hat\mu$:
\begin{equation}
0=-\frac{\partial\Psi}{\partial\hat\mu}=\lambda^2\left(-\frac{\partial\hat{\Psi}}
{\partial\hat\mu}-\frac{\hat{\mu}}{\pi}\right)
=\lambda^2\left(\hat\rho-\frac{\hat{\mu}}{\pi}\right)
\end{equation}
So the $q$ minimization implies
\begin{equation}
\pi\hat{\rho}=\hat{\mu}
\label{q_min1}
\end{equation}
Recalling (\ref{density_trans}) and (\ref{Z2}), this means that after minimizing with respect to the phase parameter $q$, the (period averaged) number
 density is simply proportional to the chemical potential:
\begin{equation}
\rho=\frac{\mu}{\pi}
\label{q_min2}
\end{equation}
This remarkable fact is independent of the form of the (complex) condensate, and simply follows from the 
transformation property (\ref{trans}) of the BdG Hamiltonian and its effect on the renormalized grand 
potential, as reflected in (\ref{psi_trans}). Note, of course, that such a relation between $\rho$ and $\mu$ {\it does  not arise} in the ${\rm GN}_2$ model, where the condensate is real and there is no phase invariance parameter $q$.

\subsection{Implications for Minimization of the Grand Potential $\Psi$ with respect to the scale 
parameter $\lambda$}

From (\ref{psi_trans}), it follows that $\Psi$ depends on the scale $\lambda$ explicitly, and also 
implicitly though the dependence of $\hat{\Psi}\equiv\Psi[\hat{\Delta}; \hat{T}, \hat{\mu}]$ on 
$\hat{T}=T/\lambda$, and on $\hat{\mu}=(\mu-q)/\lambda$. Thus we can write
\begin{eqnarray}
\frac{\partial\Psi}{\partial \lambda}&=&2\lambda\left(\hat{\Psi}+\frac{Z}{2\pi}\ln\lambda\right)+
\frac{Z\lambda}{2\pi}+\lambda^2\left(-\frac{\hat T}{\lambda}\right)\frac{\partial \hat\Psi }
{\partial\hat T}+\lambda^2\left(-\frac{\hat\mu}{\lambda}\right)\frac{\partial \hat\Psi }{\partial\hat \mu}
\nonumber \\
&&=2\lambda\left(\hat{\Psi}+\frac{Z}{2\pi}\ln\lambda+\frac{Z}{4\pi}+\frac{1}{2}\hat T\hat s+\frac{1}{2}
 \hat\mu\hat\rho\right)
\end{eqnarray}
Since $\hat\Psi=\hat u-\hat T\hat s-\hat\mu\hat\rho$, we can express the minimization condition 
$\frac{\partial\Psi}{\partial \lambda}=0$ in terms of the free energy as:
\begin{equation}
\hat{u}=-\frac{Z}{4\pi}-\frac{Z}{2\pi}\ln \lambda +\frac{1}{2} \hat\mu\hat\rho+\frac{1}{2}\hat T\hat s
\label{lambda_min1}
\end{equation}
If we impose also the condition (\ref{q_min1}) arising from the minimization with respect to the phase parameter $q$, 
we obtain the condition
\begin{equation}
\hat{u}=-\frac{Z}{4\pi}-\frac{Z}{2\pi}\ln \lambda +\frac{\hat{\mu}^2}{2\pi} +\frac{1}{2}\hat T\hat s
\label{lambda_q_min1}
\end{equation}
Alternatively, we can express these conditions in terms of the thermodynamic quantities for the general
 condensate $\Delta(x)$ in (\ref{general-delta}). Without using the condition
 (\ref{q_min2}) arising from the $q$ minimization, the $\lambda$ minimization condition (\ref{lambda_min1})
 can be written as 
\begin{equation}
u=-\frac{Z \lambda^2}{4\pi} +\frac{1}{2}\mu\,\rho+\frac{1}{2} T s +\frac{q}{2}\left(\rho-\frac{\mu}{\pi}\right) 
\label{lambda_min2}
\end{equation}
After imposing the condition (\ref{q_min2}) arising from the $q$ minimization, the last term vanishes 
and we obtain
\begin{equation}
u=-\frac{Z \lambda^2}{4\pi} +\frac{\mu^2}{2\pi}+\frac{1}{2} T s 
\label{lambda_q_min2}
\end{equation}
These conditions must hold for any form of the condensate $\Delta(x)$, and will prove very useful in 
studying the phase diagram of both the ${\rm NJL}_2$ and ${\rm GN}_2$ models.

\subsection{Transformation Property of the Consistency Condition}

The final technical ingredient before studying the thermodynamics is the effect of the transformation 
(\ref{trans}) on the consistency condition (\ref{cc}). Note that the 
consistency condition (\ref{cc}) must be satisfied also at finite $T$ and $\mu$, for the gap equation to hold.
Thus, the energy trace involves the thermodynamical Fermi factor, as in (\ref{cc}).
As with the grand potential, density, entropy and free energy, it is useful to express the consistency 
condition in terms of the condensate $\hat{\Delta}(x)$ obtained by setting the scale $\lambda=1$, and 
phase $q=0$. All we need to know is the effect of the transformation (\ref{energy-trans}) on ${\mathcal N}(E)$. 
From the form of (\ref{n-njl}) it is clear that
\begin{equation}
{\mathcal N}(E)=
\frac{1}{\lambda^2}\,\hat{{\mathcal N}}\left(\frac{E-q}{\lambda}\right)
\end{equation}
where
\begin{equation}
\hat{{\mathcal N}}(E)\equiv \frac{i}{4\sqrt{(E^2-1)(E-\hat{E}_2)(E-\hat{E}_3)}}
\end{equation}
Hence we can write the consistency condition as
\begin{eqnarray}
\int\frac{dE}{2\pi}\frac{\hat{{\mathcal N}}(E)}{1+e^{\hat\beta(E-\hat\mu)}}=0
\label{cc2}
\end{eqnarray}
Note that this integral is finite, even at $T=0$, and no 
renormalization is required. The effect of this condition is to express one of the four parameters 
$\lambda$, $q$, $\theta$ and $\nu$, in terms of the others, in a manner depending on $T$ and $\mu$.

\section{Thermodynamics of The Spiral Condensate}
\label{sec:spiral}

Before studying the general twisted kink crystal condensate, we  investigate the thermodynamics of the
 special case of the spiral condensate:
\begin{equation}
\Delta(x)=\lambda\, e^{2iqx}
\label{spiral}
\end{equation}
For this condensate, $\hat{\Delta}(x)=1$, (i.e., the vacuum fermion mass in our units), and so the thermodynamics
is simply that of a constant condensate
 of unit magnitude. The fermion spectrum is now symmetric about $0$, and so we can immediately write an
 expression for the corresponding grand potential $\hat{\Psi}$: 
\begin{eqnarray}
\hat{\Psi}
=-\frac{1}{4\pi}-\frac{\hat{T}}{\pi}\int_{1}^{\infty}dE\frac{E}{\sqrt{E^2-1}}
\ln((1+e^{-\hat{\beta}(E-\hat{\mu})})(1+e^{-\hat{\beta}(E+\hat{\mu})}))
\label{spiral-grand}
\end{eqnarray}
The full grand potential $\Psi$ for the spiral condensate (\ref{spiral}) is then obtained using
(\ref{psi_trans}).  Next we minimize the full grand potential $\Psi$ with respect to q and $\lambda$.

\subsection{Minimization with respect to the phase parameter $q$}

At $T=0$, we see from (\ref{spiral-grand}) that $\hat{\Psi}=-\frac{1}{4\pi}$, independent of $\hat{\mu}$, so that $\hat{\rho}=0$. Therefore, the condition (\ref{q_min1}), arising from the  minimization with respect to $q$, implies that $\hat{\mu}=0$ at $T=0$. In other words,
 $q=\mu$, so that the chemical potential lies at the center of the gap in the single-particle fermionic 
spectrum. With $q=\mu$, the spiral condensate (\ref{spiral}) is the ``chiral spiral'' solution proposed 
in  \cite{schon}. At nonzero temperature, the 
 $q$ minimization condition (\ref{q_min1}) can  be written explicitly as
\begin{eqnarray}
\hat{\mu}=\pi\hat{\rho}&=&\pi\frac{\partial\hat{\Psi}}{\partial \hat{\mu}}\nonumber\\
&&=2 \sinh(\hat{\beta}\hat{\mu})
\int_{1}^{\infty}dE\frac{E}{\sqrt{E^2-1}}\frac{e^{\hat{\beta}E}}{(1+e^{\hat{\beta}(E-\hat{\mu})})
(1+e^{\hat{\beta}(E+\hat{\mu})})} 
\label{q_min_spiral}
\end{eqnarray}
At low temperatures, $T\ll1$,  the main contribution to the energy integrals in (\ref{q_min_spiral}) 
comes from near the upper band edge $E=1$. So we approximate the density of states as:
\begin{equation}
\hat{\rho}(E)\approx\frac{1}{\sqrt{2}\sqrt{E-1}}
\end{equation}  
and (\ref{q_min_spiral}) becomes
\begin{eqnarray}
\hat{\mu}&\approx&\sqrt{2}\sinh(\hat{\beta}\hat{\mu})
\int_{1}^{\infty}dE\frac{1}{\sqrt{E-1}}e^{\hat{\beta}E}e^{-\hat{\beta}(E-\hat{\mu})}e^{-\hat{\beta}
(E+\hat{\mu})}=\sqrt{\frac{2\pi T}{\lambda}}\,e^{-\hat{\beta}}\,\sinh(\hat{\beta}\hat{\mu})
\end{eqnarray}
This also requires  $\hat{\mu}$=0, leading again to the chiral spiral solution with $q=\mu$. Indeed, it
 is easy to verify numerically that the finite temperature equation (\ref{q_min_spiral}) has a solution only at $\hat{\mu}=0$, for all temperature 
$T$.  Another argument in favor of $\mu=q$ at all temperatures is that instead of minimizing with respect to $q$,
we can minimize with respect to $\hat{\mu}$. Since $\hat{\Psi}$ is symmetric
under $\hat{\mu} \to - \hat{\mu}$, there must be a stationary point at $\hat{\mu}=0$, i.e., $\mu=q$. That it is a minimum can easily be seen by looking at the sign of the 2nd derivative (Taylor expansion of the integrand). Other minima (which could only come in pairs) are ruled out numerically.

Thus, we conclude that the minimization of the grand potential with respect to the phase parameter $q$ 
leads to $q=\mu$ for all temperature $T$, so $\mu$ always lies at the center of the gap. As should be  clear from this discussion, this fact can be traced directly to the phase transformation symmetry in (\ref{trans}).

Another immediate consequence of $q=\mu$ is that the grand potential  for the chiral spiral has a simple
$\mu$ dependence. This follows because (\ref{q_min1}) with $\hat{\mu}=0$ implies that $\hat{\Psi}$ is
 independent of  the chemical potential $\mu$. Indeed, when $\hat{\mu}=0$, the grand potential (\ref{spiral-grand})}can then be written as
\begin{eqnarray}
\hat{\Psi}=-\frac{1}{4\pi}-\frac{2\hat{T}}{\pi}\int_{1}^{\infty}dE\frac{E}{\sqrt{E^2-1}}
\ln(1+e^{-\hat{\beta}E})
\label{spiral-grand-2}
\end{eqnarray}
Then the general relation (\ref{psi_trans}) implies that for the chiral spiral condensate the full grand potential is
\begin{equation}
\Psi=\frac{\lambda^2}{4 \pi}\left(\ln\lambda^2-1\right)-\frac{2\lambda T}{\pi}\int_{1}^{\infty}dE
\frac{E}{\sqrt{E^2-1}}\ln(1+e^{-\hat{\beta}E})-\frac{\mu^2}{2\pi}
\label{Psi_after_q_min}
\end{equation}
Thus, the grand potential  for the chiral spiral has a simple 
$\mu$ dependence, and it is clear that $\rho=-\partial \Psi/\partial\mu=\mu/\pi$.

\subsection{Minimization with respect to the scale parameter $\lambda$}

From (\ref{Psi_after_q_min}) it also follows that the scale parameter $\lambda$ is determined only by $T$, 
independent of the chemical potential $\mu$.
Indeed, minimizing (\ref{Psi_after_q_min}) with respect to $\lambda$, we obtain the equation for the
 thermal mass scale $\lambda(T)$:
\begin{eqnarray}
0
=\lambda\frac{\ln\lambda}{\pi}-\frac{2T}{\pi}\int_{1}^{\infty}dE\frac{E}{\sqrt{E^2-1}}
\ln(1+e^{-\hat{\beta}E})+\lambda\frac{2}{\pi}\int_{1}^{\infty}dE\frac{E^2}{\sqrt{E^2-1}}
\frac{1}{1+e^{\hat{\beta}E}}
\label{lambda_min}
\end{eqnarray}
It is a simple exercise to show that this is equivalent to the general $\lambda$ minimization 
condition (\ref{lambda_q_min2}), expressed in terms of the entropy and the free energy. Note that this equation does not involve the chemical potential $\mu$,
 so the thermal mass scale $\lambda(T)$ must be independent of $\mu$.

At $T=0$, (\ref{lambda_min}) reduces to $\lambda \ln\lambda =0$, which implies that 
$\lambda(T=0)$=1, 
and the grand potential is simply
\begin{equation}
\Psi_{\rm min}^{T=0}=-\frac{1}{4\pi}-\frac{\mu^2}{2\pi}
\end{equation}
For small but nonzero temperature, the scale parameter $\lambda$ receives an exponentially small
 finite $T$ correction, found by approximating the energy integrals in  (\ref{lambda_min}):
\begin{eqnarray}
\lambda(T)\sim 1-\sqrt{2\pi T}e^{-1/T} \qquad , \qquad T\ll 1
\label{lambda-correction}
\end{eqnarray}
Applying  the same approximation to the minimized grand potential in (\ref{Psi_after_q_min})
 we find the leading small $T$ correction to the grand potential:
\begin{eqnarray}
\Psi_{\rm min}^{T\ll 1}&\sim &\frac{\lambda^2}{4 \pi}(\ln\lambda^2-1)-\sqrt{\frac{2T^3}{\pi}}e^{-1/T}-
\frac{\mu^2}{2\pi} \nonumber\\
&\sim& -\frac{1}{4\pi}-\frac{\mu^2}{2\pi}-\sqrt{\frac{2T^3}{\pi}}e^{-1/T}
\label{cs-low-T}
\end{eqnarray}
For general $T$, the temperature dependent mass scale $\lambda(T)$ can be obtained numerically from 
(\ref{lambda_min}). The scale $\lambda(T)$ decreases monotonically from the value $\lambda=1$ at $T=0$, and vanishes at a 
critical temperature
\begin{eqnarray}
T_c=\frac{e^\gamma}{\pi}\approx 0.566933
\label{tc}
\end{eqnarray}
At this temperature, $T=T_c$, the system undergoes a phase transition to a massless phase. Interestingly, 
this phase transition is independent of the chemical potential $\mu$, as follows from the fact that 
$\lambda(T)$ is independent of $\mu$. We can trace this fact directly to the simple form 
 (\ref{Psi_after_q_min}) of the grand potential for the chiral spiral condensate, after minimization 
with respect to the scale parameter $q$.

Just below $T_c$ the dependence of $\lambda$ on $T$ is nonanalytic, as can be seen from the following argument. After integrating by parts the second integral in (\ref{lambda_min}),  and expanding the Fermi factor we obtain
\begin{equation}
0=\ln\lambda+2\sum_{n=1}^\infty (-1)^{n+1}\int_1^\infty dE \frac{1}{\sqrt{E^2-1}}e^{-n \hat\beta E}
=\ln\lambda+2\sum_{n=1}^\infty (-1)^{n+1}K_0( n \hat\beta)
\end{equation}
where $K_0(x)$ is the modified Bessel function. To obtain the critical exponents near the phase transition, we analyze this equation near small values of $\lambda$. Since $\hat\beta=\lambda/T$ we expand the Bessel functions around zero:
\begin{equation}
0=\ln\lambda+2\sum_{n=1}^\infty (-1)^{n+1} \left[-\ln\left(\frac{n \lambda}{2 T}\right) -\gamma -\frac{n ^2 \lambda^2}{4 T^2}\left(\ln\left(\frac{n \lambda}{2 T}\right)+\gamma-1\right) \right]
\end{equation}
Here $\gamma$ is Euler's constant.  The $n$ sums can be evaluated in terms of the Riemann zeta function, leading to
\begin{equation}
0=\ln(T \pi)-\gamma-\frac{14 \zeta^\prime(-2) \lambda^2}{4 T^2}
\end{equation}
In particular, at the phase transition where $\lambda=0$,  the critical temperature is found to be $T_c=e^\gamma /\pi$, and to the leading order in $(T_c-T)$, for $T<T_c$:
\begin{equation}
\lambda(T)=\sqrt{\frac{2T_c}{-7\zeta^\prime(-2)}} \sqrt{T_c-T}+\dots \approx 3.06\, \sqrt{T_c(T_c-T)}+\dots
\end{equation} 

\begin{figure}[h]
\includegraphics[scale=.6]{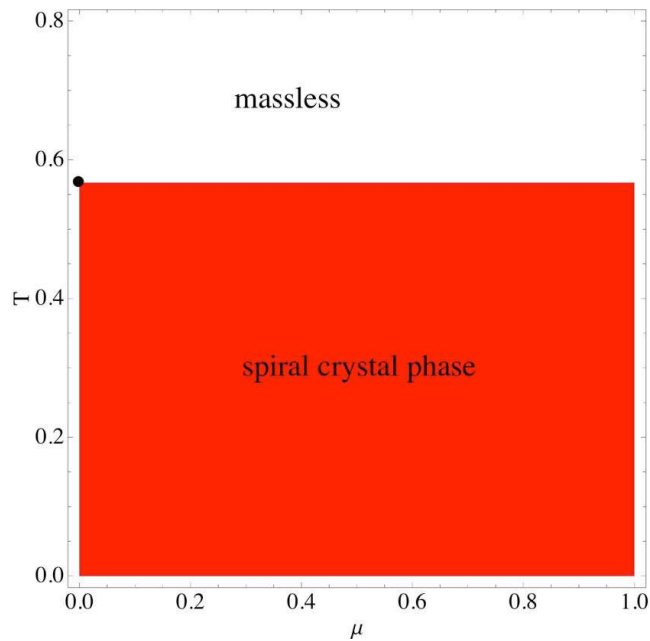}
\caption{The phase diagram of the ${\rm NJL}_2$ model. The tricritical point is marked at $\mu_{\rm tc}=0$ and $T_{\rm tc}=e^\gamma/\pi\approx 0.5669$. Below $T_c$ the condensate has the form of the spiral condensate (\ref{spiral}), with $q=\mu$.  }
\label{fig5}
\end{figure}

Thus, for the spiral condensate (\ref{spiral}), the thermodynamic phase diagram is given in Figure \ref{fig5}, showing the phase transition at $T=T_c$, independent of $\mu$. After minimizing the grand potential we learn that in the region $T<T_c$, the pitch angle $q$ of the spiral condensate is directly proportional to the chemical potential, $q=\mu$, independent of $T$, while the amplitude $\lambda(T)$ is just a function of temperature [vanishing at $T_c$], independent of $\mu$.

\section{Thermodynamics of the Twisted Crystal Condensate}
\label{sec:twisted}

In this section we develop  results for the thermodynamics of the ${\rm NJL}_2$ model with the twisted
kink crystal condensate, that is the general solution of the inhomogeneous gap equation. Recall that the twisted kink crystal condensate (\ref{complex-kink-crystal}) is characterized by 4 parameters: the scale parameter  $\lambda$, the phase parameter $q$,  an angular parameter $\theta$, and the elliptic parameter $\nu$. 
 We first consider the situation analytically at $T=0$, then at nonzero $T$.

\subsection{Twisted Kink Crystal at $T= 0$}

At $T=0$ there are some significant simplifications. First of all, the Fermi factor becomes a step function that
 acts as a cutoff of the energy integrals.
Thus, in the consistency condition (\ref{cc2}), we use 
\begin{equation}
\frac{1}{1+e^{\hat\beta(E-\hat\mu)}}\rightarrow \Theta(\hat\mu - E)
\end{equation}
If we assume that $\hat\mu$ is in the upper gap, then both the lower continuum and the bound band are 
completely filled. Thus the consistency condition (\ref{cc2}) reads (for details, see \cite{bd2})
\begin{equation}
0=\int_{-\infty}^{-1}dE\, {\mathcal N}(E)+\int_{E_2}^{E_3}dE\, {\mathcal N}(E)=\frac{1}{2A}\left(\frac{\theta}{4}
-{\bf K'}\right)
\end{equation}
Therefore, the consistency condition forces $\theta=4{\bf K}^\prime$, which is precisely the spiral 
condensate case. In this limit,  the band shrinks and joins the negative energy continuum, leaving 
just the single-gap spectrum of the  spiral condensate. Then the $q$ minimization leads to $q=\mu$ 
as described in the previous section, and we find the preferred condensate to be the chiral spiral. 
Similarly, if we assume that $\hat\mu$ is in the lower gap, then we get $\theta=0$, that is also the 
spiral condensate limit; in this limit the bound band joins to the positive energy continuum. Once again, 
minimization with respect to the phase parameter $q$ leads to $q=\mu$, so the condensate is the chiral spiral. 

The only other possibility is $\hat{\mu}$ lying inside the bound band. In this case the consistency 
condition (\ref{cc}) leads to an expression for the Fermi energy:
\begin{eqnarray}
\hat{E}_F= \hat{E}_F(\theta, \nu)={\rm nc}\left(i\theta/2; \nu\right)
\label{fermi}
\end{eqnarray}
\begin{figure}[h]
\includegraphics[scale=0.6]{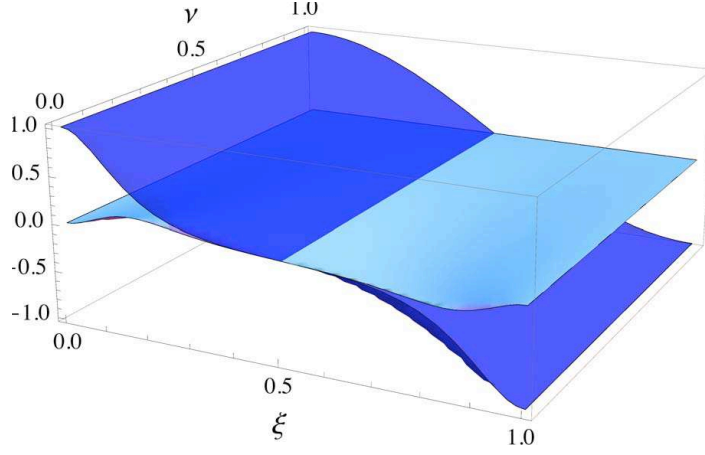}
\caption{Plot of $\pi \hat{\rho}(\xi, \nu)$ [light blue surface] and  $\hat{E}_F(\xi, \nu)$ [darker blue surface], as functions of $\nu$ and $\xi$, where $\theta\equiv 4{\bf K}^\prime \xi$. The surfaces intersect at $\xi=1/2$, which means $\theta=2{\bf K}^\prime$.} 
\label{fig6}
\end{figure}

On the other hand, at $T=0$, this Fermi energy is simply the chemical potential. Minimization with 
respect to the phase parameter $q$ leads to the relation $\hat{\mu}=\pi\,\hat{\rho}$. We can evaluate 
the density obtained by filling up to the Fermi energy $\hat{E}_F(\theta, \mu)$:
\begin{eqnarray}
\hat{\rho}= \hat{\rho}(\theta, \nu)=\frac{1}{\pi}\frac{1}{{\rm cn}(i\theta/2; \nu)}\left(\frac{{\rm cn}(i\theta/2;
 \nu)-{\rm dn}(i\theta/2; \nu)}{{\rm cn}(i\theta/2; \nu)+{\rm dn}(i\theta/2; \nu)}\right)
\label{density}
\end{eqnarray}
The simultaneous solution of these two conditions, namely $\hat{E}_F=\pi\hat{\rho}$, has the unique 
solution $\theta=2{\bf K}^\prime$, for all $\nu$,  as can be seen from Figure \ref{fig6}. Evaluating the free energy for this solution we obtain
 the following function of the remaining parameters $\lambda$ and $\nu$:
\begin{eqnarray}
{\mathcal E} = {\mathcal E}(\lambda, \nu) = \frac{2\lambda^2}{\pi(1+\sqrt{\nu})^2}\left[\frac{\nu}{2}+
\left(1-\frac{{\bf E}(\nu)}{{\bf K}(\nu)}\right)\left(\ln \left(\frac{2\lambda}{1+\sqrt{\nu}}\right)-1\right)
\right]+\frac{\mu^2}{2\pi}
\end{eqnarray}
Minimizing ${\mathcal E}(\lambda, \nu)$ with respect to $\nu$, we find that we are forced to $\nu=1$, 
which means
\begin{eqnarray}
{\mathcal E}(\lambda, \nu=1) = \frac{\lambda^2}{4\pi}\left(\ln \lambda^2-1\right)+\frac{\mu^2}{2\pi}
\end{eqnarray}
from which we recognize the $T=0$ grand potential (\ref{Psi_after_q_min}) of the chiral spiral solution. Thus, once again, the minimization  forces us to the chiral spiral condensate solution, at $T=0$.

\subsection{Twisted Kink Crystal at $0<T\ll1 $}

The minimization  of the grand potential at $T=0$ shows that the preferred twisted kink crystal configuration is the chiral spiral, with the chemical 
potential sitting in the middle of the gap (i.e $\hat{\mu}=(\mu-q)/\lambda=0$). For this solution, the 
angular parameter $\theta$ takes the values $0$ or $4{\bf K}^\prime$. We will consider the latter case 
[a similar argument applies for the other choice]. 
Now consider the stability of this chiral spiral  for $T$ nonzero but small. If we change $T$ slightly away
 from 0, then the consistency condition that sets the angular parameter $\theta=4{\bf K}^\prime$, will 
also change slightly, and the preferred value of $\theta$ will shift away from $4{\bf K}^\prime$.
We write $\theta=4{\bf K}^\prime-4\epsilon$, where $\epsilon\ll \hat{T}\ll 1$.
This changes the single-particle spectrum by producing a very narrow band very close to the lower band edge $\hat{E}_1=-1$. With $\theta=4{\bf K}^\prime-4\epsilon$, the band edges (\ref{bare_band_edges}) and the averaged amplitude (\ref{zed}) take the form:
\begin{eqnarray}
\hat{E}_2(\theta,\nu)&\approx&-1+2\nu \epsilon^2+\dots
 \nonumber \\
\hat{E}_3(\theta,\nu)&\approx&-1+2 \epsilon^2+\dots
 \nonumber \\
Z(\theta,\nu)&\approx&1-h(\nu) \epsilon^2+\dots
\label{edges_epsilon}
\end{eqnarray}
where $h(\nu)=2 \nu-2 +4 {\bf E}(\nu)/{\bf K}(\nu)$.  We now calculate the small $T$ correction to the grand potential $\hat\Psi$ for this configuration. As usual, we split the grand potential into zero temperature and finite temperature parts. For small $T$, we use the fact that  $\ln(1+e^{-|\hat\beta(E-\hat\mu)|}) \approx e^{-|\hat\beta(E-\hat\mu)|} \approx e^{-\hat\beta |E|}$ (recall that $\hat\mu=0$ for $T=0$). 
\begin{eqnarray}
\hat\Psi &\approx& \int _{-\infty}^{-1}dE \hat\rho(E)(E-\hat\mu)+ \int _{E_2}^{E_3}dE \hat\rho(E)(E-\hat\mu)-T\left( \int _{-\infty}^{-1}dE\, \hat\rho(E)e^{\hat\beta E}+ \int _{1}^{\infty}dE\, \hat\rho(E)e^{-\hat\beta E}+\int _{E_2}^{E_3}dE \,\bar\rho(E)e^{\hat\beta E}\right)
\nonumber \\
&=&<\hat \Psi>_{T=0}-Te^{-\hat\beta}\left( \int _{0}^{\infty}dx\, \hat\rho(-x-1)e^{-\hat\beta x}+ \int _{0}^{\infty}dx\, \hat\rho(x+1)e^{-\hat\beta x}+\int _{2\nu\epsilon^2}^{2\epsilon^2}dx\, \hat\rho(x-1)e^{\hat\beta x}\right)
\end{eqnarray}
In the small $T$ limit, the continuum integrals are dominated by the region $x \approx 0$ (i.e near the band edges). 
The spectral function around the band edges has the behavior:
\begin{eqnarray}
\int_0^{\infty} dx\, \hat{\rho}(-x-1) e^{-\hat{\beta} x} &=& \sqrt{\frac{T}{2\pi}} - 
\epsilon\, \frac{1}{{\bf K}(\sqrt{\nu})}\\
\int_0^{\infty} dx\,\hat\rho(x+1) e^{-\hat{\beta} x} &\approx&   \sqrt{\frac{T}{2\pi}}+ O(\epsilon^2)
\end{eqnarray}
Note that the lower continuum leads to an O($\epsilon$) correction, while the upper continuum leads to an O($\epsilon^2$) correction.
The band integral also leads to an O($\epsilon$) correction. This can be seen by 
changing the integration variable $x\rightarrow\epsilon^2 x$:
\begin{eqnarray}
\int _{2\nu\epsilon^2}^{2\epsilon^2}dE\, \hat\rho(x-1)e^{\hat\beta x}&=&\epsilon\int _{2\nu}^{2}dx\, e^{\hat\beta x \epsilon^2} \frac{ -x-(1+\nu)-h(\nu)/2}{\pi \sqrt{x(x-2\nu)(2-x)}} \approx \epsilon\int _{2\nu}^{2}dx\, \frac{ 2-x-2{\bf E}(\nu)/{\bf K}(\nu)}{\pi \sqrt{x(x-2\nu)(2-x)}} \equiv \epsilon\, L(\nu)
\end{eqnarray}
An important observation is that this function  $L(\nu)<0$ for all $\nu \in [0,1]$.

Finally, we use the general transformation of the grand potential (\ref{psi_trans}) to deduce the grand potential for the twisted kink crystal.
Since after minimization, $\hat\psi_{T=0}=-1/(4\pi)$, the  small $T$ correction to $\lambda$ does not contribute to the full grand potential $\Psi$ (as in (\ref{cs-low-T})), and the full grand potential is found to be:
\begin{eqnarray}
\Psi\approx \left(-\frac{1}{4 \pi}-\frac{\mu^2}{2\pi} -\sqrt{\frac{2 T^3}{\pi}}e^{-1/T}\right)+\epsilon\,\left( \frac{1}{{\bf K}(\sqrt{\nu})}- L(\nu)\right) T e^{-1/T}+\dots
\end{eqnarray}
The first  term in parentheses is just the low $T$ grand potential for the chiral spiral, as in (\ref{cs-low-T}).
Since $L(\nu)$ is always negative, and ${\bf K}(\sqrt{\nu})$ is always positive, we see that the system is unstable with respect to the 
opening of a  gap near the lower continuum edge. In other words, at small $T$ the minimization of the grand potential reduces the general 
twisted kink crystal condensate to the chiral spiral condensate, just as at $T=0$.

\subsection{Numerical results for the thermodynamics of the twisted kink crystal condensate}

The previous two sections have shown that at $T=0$ and for small $T$, the chiral spiral condensate is the thermodynamically preferred form of the more general twisted kink crystal condensate. We have also checked this conclusion numerically at various locations on the phase diagram, and we find that throughout the phase diagram the chiral spiral is the thermodynamically preferred limit of the general twisted kink crystal solution
of the inhomogeneous gap equation. 

\section{Thermodynamics of the real kink crystal and the ${\rm GN}_2$ model}
\label{sec:gn}

The phase diagram of the ${\rm GN}_2$ model, which has just a {\it discrete} chiral symmetry instead 
of the {\it continuous} chiral symmetry  of the ${\rm NJL}_2$ model, is  by now well understood \cite{thies-gn,feinberg}. But we revisit
 it here briefly, with a new perspective. The analysis of this paper shows that the key to understanding 
the phase diagram of the ${\rm NJL}_2$ model is the behavior (\ref{psi_trans}) of the renormalized grand potential under 
the rescaling and shifting transformation (\ref{trans}). But in the ${\rm GN}_2$ model there is no 
pseudoscalar interaction, so the condensate is real. Thus, there is no symmetry corresponding to a phase 
rotation of the condensate. In other words, $q\equiv 0$. The angular parameter $\theta$ in the solution (\ref{complex-kink-crystal}) of the inhomogeneous gap equation is also zero, as the condensate cannot wind by an arbitrary phase as it goes through one period. Furthermore, there is no need to impose any consistency condition on the solution of the gap equation: since the pseudoscalar condensate $\Pi$ is identically zero, there is no condition arising from its variation, which means that the off-diagonal terms in (\ref{gap2}) play no role. 
\begin{figure}[h]
\includegraphics[scale=0.4]{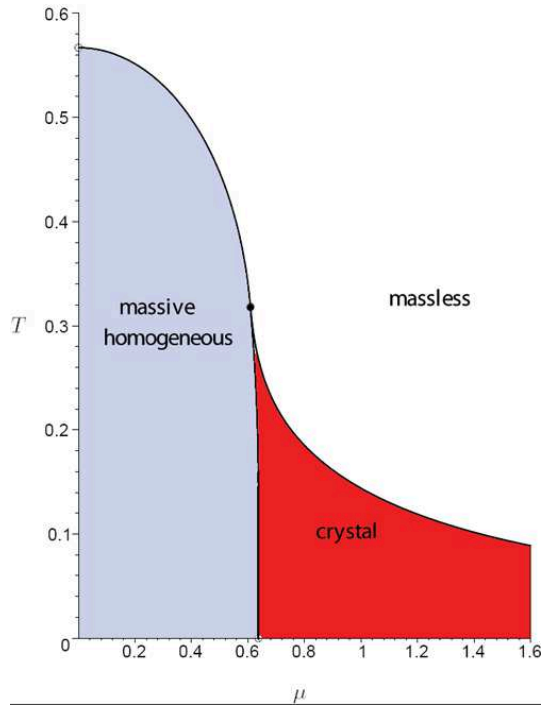}
\caption{Phase diagram of the ${\rm GN}_2$ model. The tricritical point is at $\mu_{\rm tc}=0.608$ and $T_{\rm tc}=0.318$. In the region of $\mu>2/\pi$, the massless and massive phases are separated by a crystalline phase.}
\label{fig7}
\end{figure}
Thus, the general solution (\ref{complex-kink-crystal}) of the inhomogeneous gap equation simplifies to the real kink crystal solution in (\ref{real-kink-crystal}), which depends on just two parameters, the scale $\lambda$ and the elliptic parameter $\nu$. Concerning the grand potential, the key formula is now (\ref{psi_trans}), with $q$ set to 0:
\begin{eqnarray}
\Psi_{\rm ren}[\lambda\, \hat{\Delta}(\lambda x); T, \mu]&=&\lambda^2\left(\hat{\Psi}_{\rm ren}
[\hat{\Delta}(x); T/\lambda, \mu/\lambda]+\frac{Z}{2\pi}\ln\lambda\right)
\label{psi_trans_gn}
\end{eqnarray}
The last term reflects the anomalous behavior of the grand potential under the rescaling of the condensate by $\lambda$.

\subsection{Real kink crystal at $T=0$}
From previous work \cite{thies-gn}, we know explicit expressions for the thermodynamical quantities at $T=0$ as 
functions of the elliptic parameter $\nu$.  For $\lambda=1$, the density is 
\begin{eqnarray}
\hat{\rho}=\frac{1}{2\,{\bf K}(\tilde{\nu})} \qquad , \qquad  \tilde{\nu}\equiv\frac{4\sqrt{\nu}}
{(1+\sqrt{\nu})^2}
\end{eqnarray}
The free energy is
\begin{eqnarray}
\hat{{\mathcal E}}&=& \frac{Z(\tilde{\nu})}{4\pi}\left(\ln \tilde{\nu} -1\right)+\frac{1}{2\pi}
\frac{{\bf E}(\tilde{\nu})}{{\bf K}(\tilde{\nu})}
\end{eqnarray}
where
\begin{eqnarray}
Z(\tilde{\nu})&=&2\left(1-\frac{\tilde{\nu}}{2}-\frac{{\bf E}(\tilde{\nu})}{{\bf K}(\tilde{\nu})} \right) 
\end{eqnarray}
which is just $Z(\theta, \nu)$ in (\ref{zed}), evaluated at $\theta=2{\bf K}^\prime(\nu)$. The function ${\bf E}(\tilde{\nu})$ is the complete elliptic integral of the second kind \cite{lawden}.
Thus we can write the $T=0$ grand potential as
\begin{eqnarray}
\Psi&=&{\mathcal E}-\mu\, \rho\nonumber\\
&=&\lambda^2\left(f_1(\tilde{\nu})+f_2(\tilde{\nu})\,\ln \lambda\right)-\mu\,\lambda\, f_3(\tilde{\nu})
\end{eqnarray}
where
\begin{eqnarray}
f_1(\tilde{\nu})&\equiv& \frac{Z({\tilde{\nu}})}{4\pi}\left(\ln \tilde{\nu} -1\right)+\frac{1}{2\pi}\frac{{\bf E}({\tilde{\nu}})}
{{\bf K}({\tilde{\nu}})}\\
f_2(\tilde{\nu})&\equiv& \frac{Z({\tilde{\nu}})}{2\pi}\\
f_3(\tilde{\nu})&\equiv& \frac{1}{2{\bf K}({\tilde{\nu}})}
\end{eqnarray}
Minimizing $\Psi$ with respect to $\lambda$ and $\tilde{\nu}$ leads to two equations: 
\begin{eqnarray}
\frac{\partial \Psi}{\partial \lambda}&=& 2\lambda\left(f_1+\frac{1}{2}\,f_2+f_2 \, \ln \lambda\right)-\mu\,
 f_3 =0 \nonumber\\
\frac{\partial \Psi}{\partial \tilde{\nu}}&=& \lambda\left\{\lambda\left(f_1^\prime+f_2^\prime \, 
\ln \lambda\right)-\mu\, f_3^\prime\right\} =0
\label{min}
\end{eqnarray}
Simultaneous solution of these conditions leads to a complicated-looking expression for $\ln \lambda$, 
that actually simplifies dramatically:
\begin{eqnarray}
\ln\lambda&=&\frac{f_3\, f_1^\prime-2f_3^\prime\,(f_1+f_2/2)}{2f_2\, f_3^\prime-f_3\, f_2^\prime}\nonumber\\
&=&-\frac{1}{2}\ln \tilde{\nu} 
\end{eqnarray}
In showing this remarkable reduction we use the property
\begin{eqnarray}
\frac{\partial Z}{\partial \tilde{\nu}}=\frac{(Z-\tilde{\nu})^2}{4\tilde{\nu}(1-\tilde{\nu})}
\end{eqnarray}
Inserting this result for $\lambda$ back into the minimization conditions (\ref{min}) we find the minimized 
values at $T=0$:
\begin{eqnarray}
\mu(\tilde{\nu})&=&\frac{2 {\bf E}(\tilde{\nu})}{\pi \sqrt{\tilde{\nu}}}  
\label{gn-zero-T-1}\\
\lambda(\tilde{\nu})&=&\frac{1}{\sqrt{\tilde{\nu}}}
\label{gn-zero-T-2}
\end{eqnarray}
The critical value of chemical potential, $\mu_c=\frac{2}{\pi}$, corresponding to the baryon mass \cite{gross,dhn,karsch}, is obtained 
at $\tilde{\nu}=1$, in agreement with known results  \cite{thies-gn}. 

\subsection{Real kink crystal at $T\ll 1$}

At nonzero temperature, the minimization with respect to $\lambda$ and $\tilde{\nu}$ leads to $T$ dependent 
expressions for the chemical potential and the scale factor $\lambda$, as functions of the elliptic 
parameter $\tilde{\nu}$, generalizing the $T=0$ expressions (\ref{gn-zero-T-1}, \ref{gn-zero-T-2}). This can be done numerically, 
as in \cite{thies-gn}, but here we find analytic expressions valid in the small $T$ limit.
\begin{figure}[h]
\includegraphics[scale=0.6]{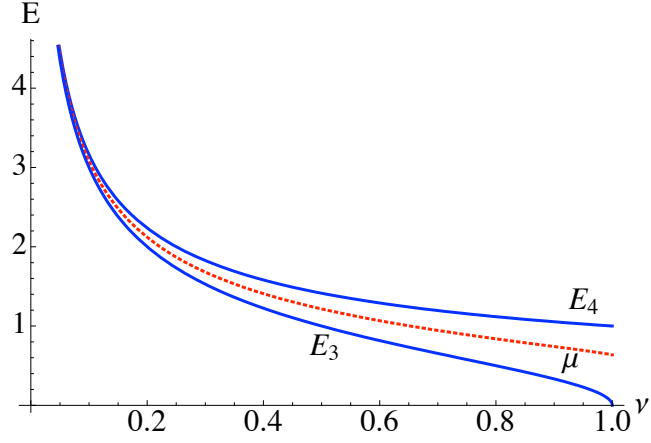}
\caption{Plot of the chemical potential [center line], and the band edge energies, as a function of the 
elliptic parameter $\tilde{\nu}$. Note that $\mu(\tilde{\nu})$ lies in the gap for all $\tilde{\nu}$, and 
moreover, it is slightly closer to the upper band edge, $E_4$, than to the lower band edge, $E_3$.}
\label{fig8}
\end{figure}
First, we note that the $T=0$ chemical potential  in (\ref{gn-zero-T-1}) lies in the upper gap [see Figure \ref{fig8}]:
\begin{eqnarray}
E_3(\tilde{\nu})=\lambda(\tilde{\nu})\sqrt{1-\tilde{\nu}} \leq \mu(\tilde{\nu})\leq 
E_4(\tilde{\nu})=\lambda(\tilde{\nu})
\end{eqnarray}
Since $\mu$ is in the gap, at small $T$ there is an exponentially small factor in the corrections to thermodynamic quantities 
going like
\begin{eqnarray}
\exp[-|\mu-{\rm nearest\,\, band\,\, edge}|/T]
\end{eqnarray}
Furthermore, for all $\tilde{\nu}$, $\mu$ is closer to $E_4$ than to $E_3$. Thus, we can write as a leading approximation
\begin{eqnarray}
\Psi&=&- T\int dE\, \rho(E)\, \ln\left(1+e^{-\beta(E-\mu)}\right)\\
&=&\Psi_{T=0}- T\int dE\, \rho(E)\, \ln\left(1+e^{-\beta |E-\mu|}\right)\\
&\sim& \Psi_{T=0}- T e^{ -\beta(E_4-\mu)} \int_{E_4}^\infty dE\, \rho(E)\, e^{-\beta (E-E_4)}
\end{eqnarray}
We expand  the spectral function in the vicinity of the nearer band edge, $E_4$:
\begin{eqnarray}
\rho(E)&=&\frac{2 E^2-(E_3^2+E_4^2)+{\lambda^2}\,Z}{2 \pi \sqrt{(E^2-E_3^2)(E^2-E_4^2)}}\\
&\sim&  \frac{1}{2\pi} \frac{E_4^2-E_3^2+{\lambda^2}\,Z}{\sqrt{2E_4(E_4^2-E_3^2)}} \frac{1}{\sqrt{E-E_4}}+O(\sqrt{E-E_4})
\end{eqnarray}
Thus,\begin{eqnarray}
\Psi\sim \Psi_{T=0}-T^{3/2}\sqrt{\frac{\lambda}{2\pi}}\, f_4(\tilde{\nu})\, e^{-\beta(E_4-\mu)}
\end{eqnarray}
where
\begin{eqnarray}
f_4(\tilde{\nu})=\frac{1}{\sqrt{\tilde{\nu}}} \,\left(1-\frac{{\bf E}(\tilde{\nu})}{{\bf K}(\tilde{\nu})}
 \right)
\end{eqnarray}
We now minimize $\Psi$ with respect to $\lambda$ and $\tilde{\nu}$, keeping the leading small $T$ corrections
 to the $T=0$ results of the previous section. We find, after some straightforward algebra, 
\begin{eqnarray}
\mu(\tilde{\nu}, T)&\sim &\frac{2 {\bf E}(\tilde{\nu})}{\pi \sqrt{\tilde{\nu}}}-
\sqrt{\frac{2 T}{\pi}}\, \frac{(1-\tilde{\nu}){\bf K}(\tilde{\nu})}{\tilde{\nu}^{3/4}}\, 
\exp\left[-\beta\left(\frac{1}{\sqrt{\tilde{\nu}}} \left(1-\frac{2 {\bf E}(\tilde{\nu})}{\pi}\right)\right)
 \right] \\
\lambda(\tilde{\nu}, T)&\sim &\frac{1}{\sqrt{\tilde{\nu}}}- \sqrt{\frac{\pi T}{2}}\, \frac{1}{\tilde{\nu}^{3/4}
}\, 
\exp\left[-\beta\left(\frac{1}{\sqrt{\tilde{\nu}}} \left(1-\frac{2 {\bf E}(\tilde{\nu})}{\pi}\right)\right)
 \right]
 \label{smallT-GN}
\end{eqnarray}
These small $T$ corrections are plotted in Figure \ref{fig9}, and are in very good agreement with the numerical results found 
in  \cite{thies-gn}, and plotted in Figure \ref{fig7}. Already these corrections indicate the existence of a crystalline phase in which the condensate 
scale $\lambda$ and the period (set by the elliptic parameter $\nu$) are dependent on {\it both}  $T$ and 
$\mu$. This is in contrast to the phase diagram of the ${\rm NJL}_2$ model, shown in Figure \ref{fig5}, where the phase transition 
 line is 
only a function of $T$,  and the scale parameter $\lambda$ is independent of the chemical potential $\mu$. With this perspective we can trace this fundamental difference in the phase diagrams
 directly to the fundamental difference between the discrete and continous chiral symmetry of the two models.
\begin{figure}[h]
\includegraphics[scale=.4]{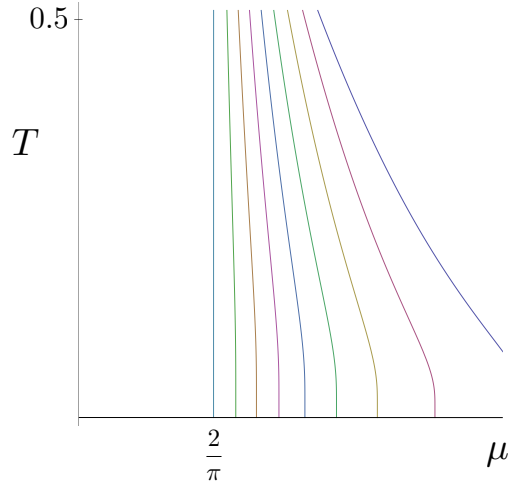}
\caption{Plots of the chemical potential, as a function of $T$, for various values of the elliptic 
parameter $\tilde{\nu}$, keeping just the leading small $T$ behavior, as in (\ref{smallT-GN}). This plot is in 
remarkable agreement with the numerical results shown in Figure \ref{fig7}.}
\label{fig9}
\end{figure}

\section{Ginzburg-Landau Analysis}
\label{sec:gl}

We complete our analysis of the phase diagram of the ${\rm NJL}_2$  and ${\rm GN}_2$ models by analyzing another region of
 the phase diagram, using the Ginzburg-Landau expansion of the grand potential $\Psi$.  In the previous sections we obtained analytic results at and near $T=0$, but the Ginzburg-Landau approach permits a certain degree of analytic information about another region of the phase diagram, in the vicinity of the tricritical point.
Expanding in powers of the condensate and its derivatives, the renormalized grand potential density may
 be expressed as:
\begin{eqnarray}
{\Psi}_{\rm GL}={\alpha}_{0}+{\alpha}_{2}|{\Delta}|^2+{\alpha}_{3}\mathrm{Im}\left({\Delta}
{\Delta}^{\prime*}\right)+{\alpha}_{4}(|{\Delta}|^4+|{\Delta}^{\prime}|^2)+{\alpha}_{5}\mathrm{Im}
\left(({\Delta}^{\prime\prime}-3|{\Delta}|^2{\Delta}){\Delta}^{\prime*}\right) \nonumber
\\ \nonumber
\\
+{\alpha}_{6}(2|{\Delta}|^6+8|{\Delta}|^2|{\Delta}^{\prime}|^2+2\mathrm{Re}{\Delta}^{\prime2}
{\Delta}^{*2}+|{\Delta}^{\prime\prime}|^2)+\dots 
\label{gl-psi}
\end{eqnarray}
The coefficients $\alpha_n(T, \mu)$ are the following functions of $T$ and $\mu$  \cite{thies-gn}:
\begin{eqnarray}
\alpha_0&=&-\frac{\pi T^2}{6}-\frac{\mu^2}{2\pi}\nonumber\\
\alpha_2&=&\frac{1}{2\pi}\left[ \ln(4\pi T)+{\rm Re}\,\psi\left(\frac{1}{2}+i\frac{\beta\mu}{2\pi}
\right)\right]\nonumber\\
\alpha_3&=& -\frac{1}{2^3\pi^2 T}{\rm Im}\,\psi^{(1)}\left(\frac{1}{2}+i\frac{\beta\mu}{2\pi}\right)
\nonumber\\
\alpha_4&=&-\frac{1}{2^6\pi^3 T^2}{\rm Re}\,\psi^{(2)}\left(\frac{1}{2}+i\frac{\beta\mu}{2\pi}\right)
\nonumber\\
\alpha_5&=&-\frac{1}{2^8\pi^4 3\, T^3}{\rm Im}\,\psi^{(3)}\left(\frac{1}{2}+i\frac{\beta\mu}{2\pi}\right)
\nonumber\\
\alpha_6&=&\frac{1}{ 2^{12}\pi^5 3\, T^4}{\rm Re}\,\psi^{(4)}\left(\frac{1}{2}+i\frac{\beta\mu}{2\pi}\right)
\label{alphas}
\end{eqnarray}

Keeping terms up to a certain order in this expansion, and inserting into the gap equation (\ref{gap1}), 
we obtain an equation (the Ginzburg-Landau equation) for the condensate $\Delta$. Remarkably, for the
 ${\rm NJL}_2$  and ${\rm GN}_2$ models, this hierarchy of equations can be solved to all orders \cite{bd2,gesztesy,correa}. If we expand up to $\alpha_2$, 
then the Ginzburg-Landau (GL) equation is simply
$
\Delta=0
$, 
so we learn nothing about the phase diagram. To this order the system appears to be just a free massless Fermi gas.
If we expand $\Psi$ up to $\alpha_3$, then the Ginzburg-Landau (GL) equation reads:
\begin{eqnarray}
\Delta^\prime-i\frac{\alpha_2}{\alpha_3 } \Delta=0\quad \Rightarrow\quad \Delta=\lambda\, 
\exp\left[i\frac{\alpha_2}{\alpha_3}\, x\right]
\end{eqnarray}
This has the form of the spiral condensate studied in Section \ref{sec:gap-spiral}. 
This spiral condensate has constant magnitude, $|\Delta|^2=\lambda^2$, and also $\frac{1}{2i}
\left(\Delta \Delta^{*\prime}-\Delta^* \Delta^{\prime}\right)=-\frac{\alpha_2}{\alpha_3}\lambda^2$, 
so that when we evaluate the grand potential on this solution, we find 
\begin{eqnarray}
\Psi_{\rm GL}=\alpha_0 \quad .
\end{eqnarray}
So the grand potential is independent of $\lambda$, and this is again no different from a free massless phase.
If we expand up to $\alpha_4$, then the Ginzburg-Landau (GL) equation is the NLSE equation:
\begin{eqnarray}
\left(\Delta^{\prime\prime}-2|\Delta|^2\Delta\right)-i\frac{\alpha_3}{\alpha_4}\Delta^\prime -\frac{\alpha_2}
{\alpha_4}\Delta=0
\label{gl-nlse}
\end{eqnarray}
The general bounded solution of this equation is the twisted kink crystal described in Section \ref{sec:gap-twisted}.

The general pattern is the following: to order $\alpha_k$, the GL equation is a differential equation of
 order $(k-2)$, and the general solution corresponds to a finite gap Dirac problem with $(k-2)$ gaps, or 
$(k-1)$ bands (including the positive and negative continuum bands). For example, the $\alpha_2$ equation 
led to $\Delta=0$, which is the free system with no gaps. The $\alpha_3$ equation leads to $\Delta=\lambda\,
 \exp\left[i\frac{\alpha_2}{\alpha_3}\, x\right]$, which has precisely one gap. The $\alpha_4$ equation,
 the NLSE (\ref{nlse}), has as its general solution a system with two gaps, as shown in the first plot of Figure \ref{fig2}. In general the solution with 
$(k-2)$ gaps requires $2(k-2)$ parameters for the solution, and these parameters can be thought of as 
labelling the band edges. Let us write
\begin{eqnarray}
{\Psi}_{\rm GL}={\alpha}_{0}(T, \mu)+\sum_{l=2}^\infty {\alpha}_{l}(T, \mu)\, J_l[\Delta, \Delta^\prime,
 \Delta^{\prime\prime}, \dots]
\end{eqnarray}
Then the GL equation to order $k$ is:
\begin{eqnarray}
\sum_{l=2}^k {\alpha}_{l}(T, \mu)\, \frac{\delta J_l[\Delta, \Delta^\prime, \Delta^{\prime\prime}, 
\dots]}{\delta \Delta^*(x)}=0
\label{novikov}
\end{eqnarray}
These equations define the AKNS hierarchy, and (\ref{novikov}) is also known as the Novikov equation. Formal
 expressions exist for their solution in terms of multi-dimensional theta functions  \cite{gesztesy,correa}, although these are cumbersome to work with beyond the twisted kink crystal solution.
The remarkable integrability properties of the AKNS hierarchy implies that the solution to the NLSE
satisfies the Novikov equations to all orders, with suitable choices of parameters, as shown also in 
\cite{bd2}.

\subsection{Ginzburg-Landau expansion for the ${\rm GN}_2$   model}

It is instructive to see how successive orders of the Ginzburg-Landau expansion reveal more and more 
about the exact phase diagram. In the ${\rm GN}_2$ system, the condensate is real, and we write it as $\Delta=\phi$, and all odd-index terms of the Ginzburg-Landau expansion vanish. The Dirac spectrum is now symmetric about $0$, and so the band edges of the finite-gap solutions come in $\pm$ pairs, as in the 2-gap case depicted in the third plot of Figure \ref{fig2}. Therefore, we only need half as many parameters at a given order to describe the solution. The grand potential density simplifies to
\begin{eqnarray}
{\Psi}_{\rm GL}={\alpha}_{0}+{\alpha}_{2}{\phi}^2+{\alpha}_{4}\left({\phi}^4+{\phi}^{\prime 2}-\frac{1}{3}
\left(\phi^2\right)^{\prime\prime}\right)+{\alpha}_{6}\left(2{\phi}^6+10{\phi}^2{\phi}^{\prime 2}+
{\phi}^{\prime\prime 2}
-\left(\phi^4+\left(\phi^\prime\right)^2-\frac{1}{5}\left(\phi^2\right)^{\prime\prime}\right)^{\prime\prime}\right)
+\dots
\end{eqnarray}
The tricritical point is defined as the point where the first two nontrivial coefficients,
 $\alpha_2(T, \mu)$ and $\alpha_4(T, \mu)$ vanish:
\begin{eqnarray}
\alpha_2(T, \mu)=\alpha_4(T, \mu)=0 \qquad \Rightarrow \qquad  T_{\rm tc}= 0.318329\quad ,
 \quad \mu_{\rm tc}= 0.608221 
\label{gn-tri}
\end{eqnarray}

\subsubsection{Ginzburg-Landau expansion to O($\alpha_4$) for the ${\rm GN}_2$   model}
\label{sec:gl-gn-4}

As mentioned above, the expansion to O($\alpha_2$) yields no information. The next nontrivial order,
 to O($\alpha_4$),  leads to the following GL equation, which is a special case of the NLSE (\ref{gl-nlse})
\begin{eqnarray}
\phi^{\prime\prime}-2\phi^3-\frac{\alpha_2}{\alpha_4}\phi=0
\end{eqnarray}
The general solution can be written as
\begin{eqnarray}
\phi=\lambda\,\sqrt{\nu}\, {\rm sn}(\lambda x; \nu) \quad \Leftrightarrow\quad
 \phi^{\prime\prime}-2\phi^3+(1+\nu)\lambda^2\,\phi=0
 \label{sn-sol}
\end{eqnarray}
with the identification of the scale parameter $\lambda$ as
\begin{eqnarray}
\lambda^2=\left(-\frac{\alpha_2}{\alpha_4}\right)\left(\frac{1}{1+\nu}\right)
\label{gna2-4}
\end{eqnarray}
Notice that in the GL approach, we get explicit expressions for the dependence of the solution's
 parameters in terms of $T$ and $\mu$. An important comment is that since $\frac{1}{1+\nu}\geq 0$,
 this expression tells us that this inhomogeneous solution only makes sense in regions of the $(T, \mu)$ 
plane where $\left(-\frac{\alpha_2}{\alpha_4}\right)\geq 0$.
Using the following identities satisfied by the solution in (\ref{sn-sol})
\begin{eqnarray}
\left(\phi^\prime\right)^2&=&\phi^4-(1+\nu)\lambda^2\phi^2+\nu\, \lambda^4 \\
\left(\phi^2\right)^{\prime\prime}&=& 6\phi^4-4(1+\nu)\lambda^2\,\phi^2+2\nu\, \lambda^4
\end{eqnarray}
we can write the grand potential density to this order as
\begin{eqnarray}
\Psi_{\rm GL}&=&{\alpha}_{0}+{\alpha}_{2}{\phi}^2+{\alpha}_{4}\left(\frac{1}{3}(1+\nu)\lambda^2 
\phi^2+\frac{1}{3}\nu\, \lambda^4\right)\nonumber\\
&=&{\alpha}_{0}+\frac{2}{3}{\alpha}_{2}\,\phi^2+\frac{\nu}{3(1+\nu)^2}\,\frac{\alpha_2^2}{{\alpha}_{4}}
\end{eqnarray}
Here we have used the above expression (\ref{gna2-4}) for $\lambda^2$.
Averaging over one period, we use $\langle \phi^2\rangle =\lambda^2(1-{\bf E}(\nu)/{\bf K}(\nu))$,
 and again using (\ref{gna2-4}) we find
\begin{eqnarray}
\langle\Psi\rangle^{\rm crystal}_{\rm GL} &=&{\alpha}_{0}+\left(-\frac{\alpha_{2}^2}{4\alpha_4}\right)
\left[\frac{4}{1+\nu}\left(\frac{2+\nu}{3(1+\nu)}-\frac{2}{3}\frac{{\bf E}(\nu)}{{\bf K}(\nu)}\right)
\right]\nonumber\\
&\equiv& {\alpha}_{0}+\left(-\frac{\alpha_{2}^2}{4\alpha_4}\right)F(\nu)
\label{gn-gl-energy-4}
\end{eqnarray}
Note that the function $F(\nu)$ is a smooth function interpolating monotonically between $F(0)=0$ and $F(1)=1$. 
We have written $\langle\Psi\rangle^{\rm crystal}_{\rm GL}$ like this in order to compare with the homogeneous 
ansatz: $\phi=\lambda$. Then 
\begin{eqnarray}
\langle\Psi\rangle^{\rm homogeneous}_{\rm GL}&=&{\alpha}_{0}+\alpha_2 \lambda^2+\alpha_4 \lambda^4
\end{eqnarray}
Minimizing with respect to $\lambda^2$, we obtain the condition $\lambda^2=-\alpha_2/(2 \alpha_4)$, and at this 
minimum
\begin{eqnarray}
\langle\Psi\rangle^{\rm homogeneous}_{\rm GL}={\alpha}_{0}+\left(-\frac{\alpha_{2}^2}{4\alpha_4}\right)
\end{eqnarray}
Therefore, we can write
\begin{eqnarray}
\langle\Psi\rangle^{\rm homogeneous}_{\rm GL}
-\langle\Psi\rangle^{\rm crystal}_{\rm GL}=\left(-\frac{\alpha_{2}^2}{4\alpha_4}
\right)\left[1-F(\nu)\right]
\end{eqnarray}
An important observation is that at the values of ${\nu}=1$ and ${\nu}=0$, the minimized grand potential
 reduces to that of the homogeneous and the massless condensates (recall ${\bf E}/{\bf K}({\nu=1})=0,
{\bf E}/{\bf K}({\nu=0}$)=1):
\begin{eqnarray}
\langle {\Psi}\rangle^{\rm crystal}_{\rm GL}({\nu}=1)&=&{\alpha}_{0}+\left(-\frac{\alpha_{2}^2}{4\alpha_4}\right)
=\langle {\Psi}\rangle^{\rm homogenous}_{\rm GL}
\nonumber \\
\langle {\Psi}\rangle^{\rm crystal}_{\rm GL}({\nu}=0)&=&{\alpha}_{0}=\langle {\Psi}\rangle^{\rm massless}_{\rm GL}
\end{eqnarray}
\begin{figure}[h]
\includegraphics[scale=0.5]{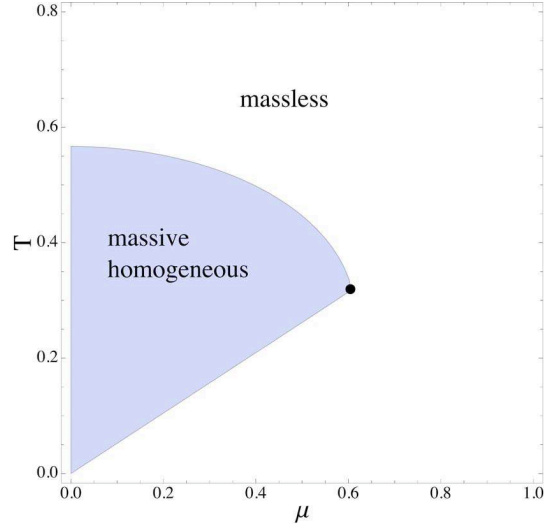}
\caption{The phase diagram of the ${\rm GN}_2$ model, based on a Ginzburg-Landau expansion to  the lowest nontrivial order: O($\alpha_4$). The blue region is the region in which the massive homogeneous condensate $\phi=\lambda$ has a lower grand potential. The white region is the region in which the homogeneous massless condensate $\phi=0$  has a lower grand potential, or where only the massless condensate exists, because $\lambda^2$ in (\ref{gna2-4}) is negative. These regions meet at the tricritical point: $T_{\rm tc}= 0.318$, $\mu_{\rm tc}= 0.608$. } 
\label{fig10}
\end{figure}
This behavior is depicted in Figure \ref{fig10}, where the grand potential of the crystal condensate lies between that of the massless and massive homogeneous phases, interpolating between them as a function of $\nu$. Minimizing with respect to $\nu$ pushes us to the massive  homogeneous phase in the blue region, but to the massless homogeneous phase in the white region. Thus, at this order of the GL expansion, even though the solution to the GL equation has the form of a crystalline condensate, the thermodynamic minimum is a constant condensate, either zero or non-zero, but always constant. 
We show in the next section that this picture changes significantly at the next order.

\subsubsection{Ginzburg-Landau expansion to O($\alpha_6$) for the ${\rm GN}_2$   model}
\label{sec:gl-gn-6}

Going to the next non-trivial order beyond the level defining the tricritical point, we expand the 
grand potential density in powers of the real condensate field $\phi$ and its derivatives [we drop the total derivative terms as these are not important for this argument]:
\begin{eqnarray}
{\Psi}_{\rm GL}={\alpha}_{0}+{\alpha}_{2}\phi^2+{\alpha}_{4}(\phi^4+\phi^{\prime 2})+
{\alpha}_{6}(2{\phi}^6+10{\phi}^2{\phi}^{\prime 2}+{\phi}^{\prime\prime 2})
\end{eqnarray}
The GL equation is  now a fourth-order equation:
\begin{eqnarray}
\left(\phi^{\prime\prime\prime\prime}-10\phi^2\phi^{\prime\prime}-10\phi(\phi^\prime)^2+
6 \phi^5\right)+\frac{\alpha_4}{\alpha_6}\left(-\phi^{\prime\prime}+2\phi^3\right)+
\frac{\alpha_2}{\alpha_6}\phi=0
\label{gl-a6}
\end{eqnarray}
The simplest solution is a homogeneous condensate, $\phi=\lambda$, with massless and massive solutions:
\begin{eqnarray}
\lambda&=&0 \qquad ({\rm massless\,\, homogeneous\,\, phase})\nonumber\\
\lambda^4+\frac{\alpha_4}{3\alpha_6}\lambda^2+\frac{\alpha_2}{6\alpha_6}&=&0\quad \Rightarrow 
\quad \lambda_\pm^2=-\frac{\alpha_4}{6\alpha_6}\left(1\pm \sqrt{1-\frac{6\alpha_2\alpha_6}
{\alpha_4^2}}\right) \quad ({\rm massive\,\, homogeneous\,\, phase})
\end{eqnarray}
The general solution to (\ref{gl-a6}) is very complicated, but we can use the inhomogeneous 
solution to the NLSE 
\begin{equation}
\phi=\lambda \sqrt{{\nu}}\,{\rm sn}(\lambda\, x,{\nu})
\label{gl-real-crystal}
\end{equation} 
A similar idea was used in an analogous condensed matter model in \cite{buzdin}.
This solution satisfies the nonlinear equations: 
\begin{eqnarray}
-\phi^{\prime\prime}+2\phi^3&=&(1+\nu)\lambda^2 \phi\nonumber\\
\left(\phi^{\prime\prime\prime\prime}-10\phi^2\phi^{\prime\prime} - 10 \phi (\phi')^2+6\phi^5\right)&=&(\nu^2+4\nu+1)
 \lambda^4 \phi
\end{eqnarray}
\begin{figure}[h]
\includegraphics[scale=1.25]{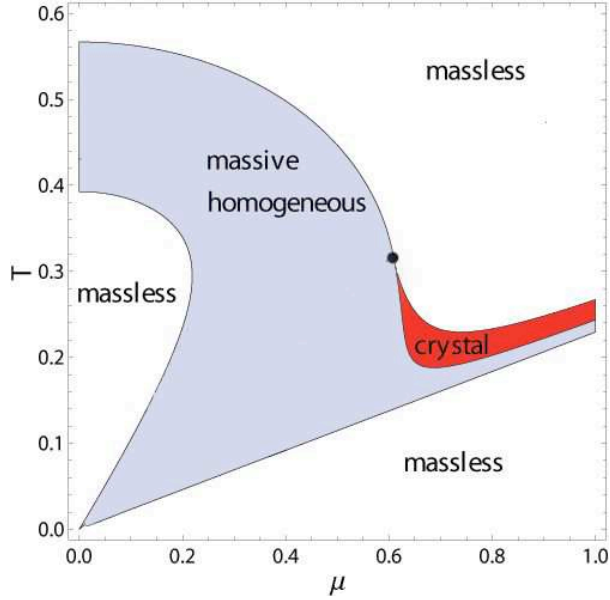}
\caption{The phase diagram of the ${\rm GN}_2$ model, based on a Ginzburg-Landau expansion to  the lowest nontrivial order: O($\alpha_6$). The blue region is the region in which the massive homogeneous condensate $\phi=\lambda$ has the lowest grand potential. The red region is the region in which the crystalline condensate (\ref{gl-real-crystal}) has the lowest grand potential. In the remaining [white] region,  the
homogeneous massless condensate $\phi=0$ has the lowest grand potential. Note that the crystal phase region begins at the tricritical point: $T_{\rm tc}= 0.318$, $\mu_{\rm tc}= 0.608$. A close-up of this region is shown in Figure \ref{fig12}.} 
\label{fig11}
\end{figure}
Thus, comparing with the GL equation (\ref{gl-a6}), we see that $\phi$ satisfies the GL equation (\ref{gl-a6}) provided we identify:
\begin{eqnarray}
\lambda^4+\frac{\nu+1}{(\nu^2+4\nu+1)}\frac{\alpha_4}{\alpha_6}\lambda^2+\frac{1}{(\nu^2+4\nu+1)}
\frac{\alpha_2}{\alpha_6}=0
\end{eqnarray}
This condition leads to two solutions
\begin{eqnarray}
\lambda_\pm^2=-\frac{\nu+1}{2(\nu^2+4\nu+1)}\frac{\alpha_4}{\alpha_6}\left(1\pm 
\left(1-\frac{4(\nu^2+4\nu+1)}{(\nu+1)^2}\frac{\alpha_2 \alpha_6}{\alpha_4^2}\right)^{1/2}\right)
\label{lambdapm}
\end{eqnarray}
Evaluated on the crystalline solution, the grand potential is
\begin{eqnarray}
\langle {\Psi}\rangle_{\rm GL}={\alpha}_{0}+\lambda^2{\alpha}_{2}\left( 1-\frac{{\bf E}}{{\bf K}} \right)
+\frac{\lambda^4{\alpha}_{4}}{3}\left(1+2{\nu}-(1+{\nu})\frac{{\bf E}}{{\bf K}}\right)
+\frac{\lambda^6{\alpha}_{6}}{5}\left(3{\nu}^2+6{\nu}+1-({\nu}^2+4{\nu}+1)\frac{{\bf E}}{{\bf K}}\right)
\label{rl_free_energy}
\end{eqnarray}
\begin{figure}[h]
\includegraphics[scale=.4]{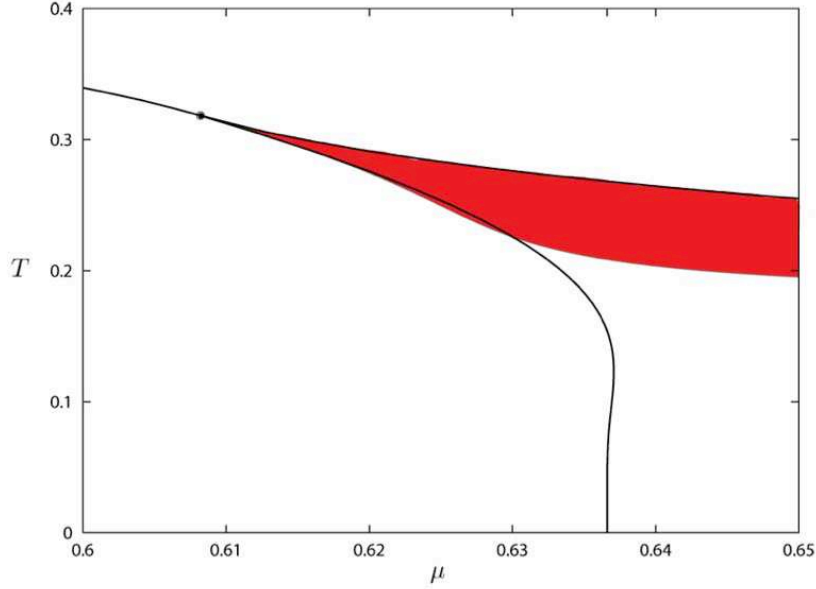}
\caption{A close-up view of the crystalline region in the phase diagram of the ${\rm GN}_2$ model, near the tricritical point, based on a Ginzburg-Landau expansion to  the lowest nontrivial order: O($\alpha_6$). The red shaded region is the crystalline region seen at this order of the GL expansion, while the solid black lines mark the edges of the true crystalline region found numerically from the exact grand potential \cite{thies-gn}. The agreement is excellent near the tricritical point and near the LOFF boundary with the massless phase.} 
\label{fig12}
\end{figure}
This is just a function of $T$ and $\mu$ (through the $\alpha$'s) and the elliptic parameter $\nu$, 
because $\lambda$ is given by the solutions in (\ref{lambdapm}). We can therefore evaluate the grand 
potential throughout the $(T, \mu)$ plane and ask where it is lower than the grand potential of the 
homogeneous phase. The result is shown in Figure \ref{fig11}, which shows the existence of a crystalline phase
 in a small region in the vicinity of the tricritical point. This is a region in which the grand potential of
 the crystalline condensate is lower than that of the massless or massive homogeneous condensate. On 
the upper edge, $\nu=0$ and the scale of the crystalline condensate vanishes as it  reduces to a 
massless phase; on the lower edge, $\nu=1$, and the period of the crystalline condensate diverges as 
it reduces to a homogeneous massive phase. The form of this region matches very well with the full crystalline region, near the tricritical point, as shown by the close-up view in Figure \ref{fig12}.
Going to higher orders of the GL expansion, this crystalline
 region grows, and eventually covers the entire region given by the exact numerics \cite{thies-gn}.

\subsection{Ginzburg-Landau expansion for the ${\rm NJL}_2$   model}
\label{sec:gl-njl}

In contrast to the ${\rm GN}_2$ model,  the ${\alpha}_{3}$ term  in (\ref{gl-psi}) is present in the GL expansion of the ${\rm NJL}_2$ model, as the condensate $\Delta$
 is complex.  The ``tricritical'' point is defined as the point where the two lowest nontrivial coefficients,
 $\alpha_2(T, \mu)$ and $\alpha_3(T, \mu)$, vanish:
\begin{eqnarray}
\alpha_2(T, \mu)=\alpha_3(T, \mu)=0 \qquad \Rightarrow \qquad  T_{\rm tc}=0.566\quad ,\quad  \mu_{\rm tc}= 0
\label{njl-tri}
\end{eqnarray}

\subsubsection{Ginzburg-Landau expansion to O($\alpha_3$) for the ${\rm NJL}_2$  model}
\label{sec:gl-njl-3}

The first nontrivial order, to O($\alpha_3$), leads to the GL equation
\begin{eqnarray}
\Delta^\prime-i\frac{\alpha_2}{\alpha_3 } \Delta=0\quad \Rightarrow\quad \Delta=\lambda\, 
\exp\left[i\frac{\alpha_2}{\alpha_3}\, x\right]
\end{eqnarray}
But for this solution, even though this condensate is crystalline, the grand potential is $\langle \Psi\rangle_{\rm GL}=\alpha_0$. Thus, the phase diagram is simply that of a massless 
phase. The only thing we learn at this level of the GL expansion is the existence of the tricritical point at $T=0.5669$ and $\mu=0$. This is analogous to the situation of the GL expansion of the ${\rm GN}_2$ model to its first nontrivial order, O($\alpha_4$), where the solution of the GL equation has a crystalline form, but this crystalline condensate does not appear in the phase diagram at that order, as discussed in Section \ref{sec:gl-gn-4}.

\subsubsection{Ginzburg-Landau expansion to O($\alpha_4$) for the ${\rm NJL}_2$  model}
\label{sec:gl-njl-4}

Going to the next non-trivial order beyond the level defining the tricritical point, namely to O($\alpha_4$),
 we obtain the GL equation of NLSE form in (\ref{gl-nlse}).
Adapting the solution in Section \ref{sec:gap-twisted}), we  can write the  general solution  as
\begin{equation}
{\Delta}=-\lambda \frac{{\sigma}(\lambda x+i{\bf K^{\prime}}-i{\theta}/2)}{{\sigma}(\lambda x+
i{\bf K^{\prime}}){\sigma}(i{\theta}/2)}\exp\left[i\lambda x(-i{\zeta(i{\theta}/2)+{q}})+i 
{\theta}{\eta}_{3}/2\right]
\label{cx_soln}
\end{equation}
which satisfies
\begin{eqnarray}
-{\Delta}^{\prime\prime}+2{\Delta}|{\Delta}|^2=-2i\, q\,\lambda\,{\Delta}^{\prime}+\lambda^2(-3{\mathcal P}
(i{\theta/2})-{q}^2){\Delta}
\end{eqnarray}
Identifying the terms with the NLSE equation we deduce ${q}=\frac{{\alpha}_{3}}{2\lambda
{\alpha}_{4}}$, and
 $\lambda$ must satisfy: 
\begin{eqnarray}
\lambda^2&=&\left( -\frac{{\alpha}_{2}}{2{\alpha}_{4}}\left[1-\frac{{\alpha}_{3}^2}{4\alpha_2
 {\alpha}_{4}}\right]\right)\left(\frac{2}{-3{\mathcal P}(i {\theta}/2)}\right)
\label{njl-a2-4}
\end{eqnarray}
Note that $(-{\mathcal P}(i {\theta}/2))\geq 0$. Thus, this inhomogeneous crystal condensate only makes sense 
in regions of the $(T, \mu)$ plane where $\left( -\frac{{\alpha}_{2}}{{\alpha}_{4}}
\left[1-\frac{{\alpha}_{3}^2}{4\alpha_2 {\alpha}_{4}}\right]\right)\geq 0$.
\begin{figure}[h]
\includegraphics[scale=0.5]{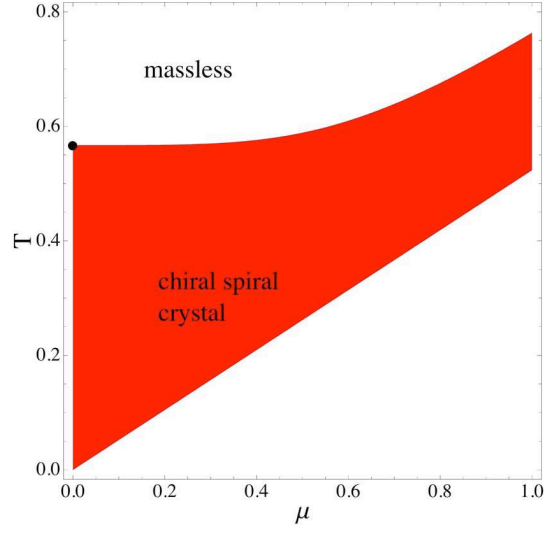}
\caption{The phase diagram of the ${\rm NJL}_2$ model, based on a Ginzburg-Landau expansion to   O($\alpha_4$). The purple region is the region in which the spiral condensate (\ref{spiral}) has a lower grand potential. The red region is the region in which the  massless condensate $\Delta=0$  has a lower grand potential, or in which only the massless condensate exists, because $\lambda^2$ in (\ref{njl-a2-4}) is negative. } 
\label{fig13}
\end{figure}

Now evaluating the averaged potential on this solution, we find
\begin{eqnarray}
\langle {\Psi}\rangle^{\rm crystal}_{\rm GL}&=&{\alpha}_{0}+\left[-\frac{{\alpha}_{2}^2}{4\alpha_4}
\left(1-\frac{{\alpha}_{3}^2}{4\alpha_2 {\alpha}_{4}} \right)^2\right]\left[\frac{4}{9}
\left(1+\frac{{\nu}^2-{\nu}+1}{9{\mathcal P}(i {\theta}/2)^2}+\frac{2}{{\mathcal P}(i 
{\theta}/2)}\frac{{\eta}}{{\bf K}}\right)
\right]
\nonumber\\
&\equiv &{\alpha}_{0}+\left[-\frac{{\alpha}_{2}^2}{4\alpha_4}\left(1-\frac{{\alpha}_{3}^2}{4\alpha_2 
{\alpha}_{4}} \right)^2\right] \,F({\nu},{\theta})
\label{njl-gl-energy-4}
\end{eqnarray}
which should be compared with the corresponding ${\rm GN}_2$ expression (\ref{gn-gl-energy-4}). [Indeed, setting 
$\theta=2{\bf K}^\prime$, and $\alpha_3=0$, we recover the ${\rm GN}_2$ formulas]. We note that 
$
0\leq F({\nu},{\theta})\leq 1
$. We now compare the averaged potential for the crystal with that obtained from a spiral ansatz: 
\begin{eqnarray}
\Delta^{\rm spiral}=\lambda\, e^{2 i q x}
\end{eqnarray}
With this spiral  ansatz we find 
\begin{eqnarray}
\langle {\Psi}\rangle^{\rm spiral}_{\rm GL}=\alpha_0+\lambda^2\alpha_2-2 q \lambda^2 \alpha_3+(\lambda^4+4 q^2 
\lambda^2) \alpha_4
\end{eqnarray}
Minimizing with respect to $q$ we find $q=\alpha_3/\alpha_4$, and further minimizing with respect to $\lambda^2$ we find 
\begin{eqnarray}
\lambda^2&=&\left( -\frac{{\alpha}_{2}}{2{\alpha}_{4}}\left[1-\frac{{\alpha}_{3}^2}{4\alpha_2 
{\alpha}_{4}}\right]\right)
\label{njl-a2-4-hom}
\end{eqnarray}
which should be compared with (\ref{njl-a2-4}). Furthermore, evaluating the averaged potential on this 
spiral condensate we find
\begin{eqnarray}
\langle {\Psi}\rangle^{\rm spiral}_{\rm GL}&=&{\alpha}_{0}+\left[-\frac{{\alpha}_{2}^2}{4\alpha_4}
\left(1-\frac{{\alpha}_{3}^2}{4\alpha_2 {\alpha}_{4}} \right)^2\right]
\label{njl-gl-energy-4-hom}
\end{eqnarray}
which should be compared with (\ref{njl-gl-energy-4}).

Now we combine this expression with the positivity condition on $\lambda^2$ in (\ref{gna2-4})
 to obtain the  phase diagram in Figure \ref{fig13}. Just as in the ${\rm GN}_2$ case, here in the in the ${\rm NJL}_2$ model, by going one step beyond the  first nontrivial order of the GL expansion [i.e., one step beyond the order  that defines the tricritical point]  we see the appearance of a crystalline phase in the phase diagram, 
in the region near the tricritical point. For the ${\rm NJL}_2$ model, the condensate of this
crystalline phase, derived from this GL approach to this order, has the form of the chiral spiral
after minimization of the grand potential. This Ginzburg-Landau analysis confirms once again that the chiral spiral is the thermodynamically 
preferred form of the  inhomogeneous condensate, in the applicable part of the phase diagram. The pattern is fairly clear: going to higher orders of the GL expansion, the crystalline region grows, and eventually covers the entire region given by the exact numerics, as shown in Figure \ref{fig5}.

\section{Conclusions}

We have used the exact  crystalline solutions to the inhomogeneous gap equation of the ${\rm NJL}_2$ model, found in \cite{bd1,bd2},  to probe the thermodynamic phase diagram of the ${\rm NJL}_2$ and ${\rm GN}_2$ models at finite density and temperature. Using a combination of exact, numerical and Ginzburg-Landau approaches, we have shown that for the ${\rm NJL}_2$ model the thermodynamically preferred condensate in the region $T<T_c$ is the helical chiral spiral of \cite{schon}. The same methods have been applied to the ${\rm GN}_2$ model, confirming previous numerical results \cite{thies-gn}. A key new idea in our analysis is the exploitation of the behavior of the grand potential under the rescaling and phase rotation transformations (\ref{trans}), which affect the renormalized grand potential as in (\ref{psi_trans}). This observation greatly facilitates the minimization of the renormalized grand potential with respect to the parameters $\lambda$ and $q$. We are also able to trace in a very explicit manner the consequences for the phase diagram of  the fact that the ${\rm GN}_2$ model has a discrete chiral symmetry, while the the ${\rm NJL}_2$ model has a continuous chiral symmetry. 
These one dimensional models are somewhat special, due to the rich integrability structure underlying their gap equation. So, we studied these models also using the Ginzburg-Landau approach, which does not necessarily rely on this integrability structure. We found that  in both the ${\rm NJL}_2$ and ${\rm GN}_2$ models the crystalline region appears at the order of the Ginzburg-Landau expansion {\it one step beyond} the first nontrivial order, which is used to identify the relevant tricritical point. It would be interesting to study this point systematically in higher dimensional models, where the search for crystalline phases is considerably more difficult \cite{buzdin,casalbuoni2,mannarelli,nickel}. It would also be interesting to study the ${\rm NJL}_2$  system on the lattice, complementing the ${\rm GN}_2$ work of \cite{karsch,deforcrand}, and the recent Monte Carlo formulations in \cite{gattringer,wolff2,klaus}. Other interesting effects include the study of an isospin chemical potential \cite{ebert}, and going beyond the leading large $N$ approximation \cite{kneur}.

\section{Acknowledgments}
GB and GD thank the DOE for support through grant DE-FG02-92ER40716, and GD thanks the  Institut f\"ur Theoretische Physik at Erlangen for support during a visit. We are grateful to F. Correa, V. Enolskii, F. Gesztesy and  D.  Nickel for helpful comments and suggestions.

\end{document}